\documentclass[pretwocolumn]{revtex4-1}%
\usepackage{graphicx}
\usepackage{amsmath}
\usepackage{ulem}
\usepackage{enumitem}
\usepackage{color}
\setcitestyle{square}
\usepackage{subfigure}

\usepackage{amssymb}

\begin{document}
\title{Fractional Edgeworth Expansion: Corrections to the Gaussian-L\'evy
Central Limit Theorem}
\author{Netanel Hazut}
\author{Shlomi Medalion}
\thanks{These two authors contributed equally}
\author{David A. Kessler}
\author{Eli Barkai}
\affiliation{Department of Physics and Institute of Nanotechnology and Advanced
Materials,
Bar-Ilan University, Ramat-Gan 52900, Israel}
\date{\today}

\begin{abstract}
{In this article we generalize the classical Edgeworth expansion for the
probability density function (PDF) of sums of a finite number of symmetric
independent identically distributed random variables with a finite variance to
PDFs with a diverging variance, which converge to a L\'evy $\alpha$-stable
density function. Our correction may be written by means of a series of
fractional derivatives of the L\'evy and the conjugate L\'evy PDFs. This series
expansion is general and applies also to the Gaussian regime. To describe the
terms in the series expansion, we introduce a new family of special functions
and briefly discuss their properties. We implement our generalization to the
distribution of the momentum for atoms undergoing
Sisyphus cooling, and show the improvement of our leading order 
approximation compared to previous approximations.
In vicinity of the transition between L\'{e}vy and Gauss behaviors, convergence
to asymptotic results slows down.}

\end{abstract}

\maketitle
\section{Introduction}
\label{Introduction} 
Statistical physics deals with systems consisting of large number of
particles. The state of these systems is generally described by their
probability density function (PDF), which enables us to determine the possible
states of the system and to calculate macroscopic quantities such as
physical average observables. Usually, Gaussian PDFs appear whenever one deals
with systems composed of a large number of particles. These PDFs
describe well systems with dynamics that is characterized by a large number of
random small events, e.g., particle motion in a liquid (Brownian
motion). However, not all systems are described by the Gaussian PDF. Many
systems are characterized by (rare) large fluctuations. These
large fluctuations give rise to a long, power-law tail in the PDF. The long tail in many cases leads to the divergence of the
second moment.

Indeed, for the two kinds of systems described above, there exist limit
theorems which give the asymptotic PDFs of the sum of random variables. The
Gaussian Central Limit Theorem (CLT) applies in the case of summation of independent,
identically distributed (iid) random variables with common PDFs characterized by
a finite variance, while a generalized
CLT \cite{feller2008introduction,metzler2000random} for
long tailed PDFs, in which case the limiting distribution is a L\'evy
distribution. However these limit theorems are valid only in the limit where the
number of random variables, $n$, goes to infinity. Hence, for many physical
systems composed of a relatively small number of particles one cannot use the
CLTs for approximating the PDF of the sum. Better approximations for finite $n$
were developed for PDFs that approach a Gaussian in the limit. Among these is
the classical Edgeworth expansion \cite{edgeworth1906generalised} which
provides asymptotic correction terms to the CLT.
Recently an improvement of the classical Edgeworth expansion was given by
Lam, et al. \cite{lam2011corrections}. This expansion generalizes the
Edgeworth result to
cases in which each of the random variables are distributed with heavy-tailed
(power-law decaying) PDFs with finite variance, but diverging higher moments.

In this article we further generalize the Edgeworth correction for cases of
random variables with diverging variance. We present correction terms for
finite $n$ for PDFs approaching the L\'{e}vy distribution. We show that our
correction is general in the sense that for PDFs for which all the
moments exist, it converges to the classical Edgeworth result, and when higher
moments diverge, to Lam et al's generalization.

In section \ref{sec::CLTandEdgeworth} we review the CLT and the
derivation of the classical Edgeworth series. In section \ref{sec:levi-edge} we
derive our generalized series and investigate the behavior of the correction
terms. In addition we present the leading term approximation and discuss the
two regimes (Gaussian and L\'evy). In section \ref{sec::ColdAtoms} we implement
our approximation to the sum of the momenta of cold atoms in an optical lattice and we
show its convergence to the exact solution (calculated numerically), and
compare it to previous approximation methods. In section \ref{sec::summary} we
summarize our results, and highlight the importance of the family of the special
functions introduced in the correction terms of our series.

\section{The CLT and the Classical Edgeworth Expansion}
\label{sec::CLTandEdgeworth}

For a set of $n$ identically independent distributed (iid) random variables,
$\{x_j\}$, with a common symmetric probability density function (PDF),
$w(x)$, with zero mean ($\mu=0$) and finite variance, $\sigma^2$, the central
limit theorem (CLT) states that for $n\rightarrow\infty$, the PDF $w_S(x)$ of
the normalized sum, $S_n \equiv \sum_{j=1}^nx_j/n^{1/2}$ is given by the
Gaussian density function:
\begin{equation}
\lim_{n\rightarrow\infty}w_S(x)= Z_\sigma(x)=\frac{1}{\sqrt{2 \pi
\sigma^2}} e^{-x^2/{2
\sigma^2}}.
\label{eq:CLT}
\end{equation}
Since for finite $n$ there are deviations from the normal density, one
might want to approximate these deviations quantitatively. A few series
expansions
for non-Gaussian densities have been suggested for this purpose, such as the
Gram-Charlier series \cite{cramer1999mathematical, kendall1977advanced} and the
Gauss-Hermite expansion \cite{blinnikov1998expansions}. The most
accurate among those is the Edgeworth expansion, since it is a true asymptotic
one \cite{juszkiewicz1995weakly, blinnikov1998expansions}. 

In order to derive the Edgeworth expansion for the density function of
the probability of the normalized sum $S_n$, we
shall introduce the characteristic function for the single variable,
$\tilde{w}(k)=\langle \exp(ikx) \rangle =
\int_{-\infty}^{\infty}w(x)\exp(ikx)dx$, and
its
logarithm, $\psi(k)=\ln \tilde{w}(k)$, so that the obtained characteristic
function for
$S_n$ can be written as $\tilde{w}_{S_n}(k) = \tilde{w}(k/ \sqrt{n})^n$ yielding
$w_{S_n}(x)$  via an
inverse Fourier transform. Alternatively, one may define
$\psi_{S_n}(k)=\ln \tilde{w}_{S_n}(k)=n \psi(k/\sqrt{n})$ and use the
inverse Fourier
transform of $\exp[\psi_{S_n}(k)]$.

In what follows we consider symmetric PDFs ($w(x)=w(-x)$).
We begin by expanding $\tilde{w}(k)$ in a power series:
\begin{equation}
\tilde{w}(k) = 1 + \sum_{j=1}^{\infty}{\frac{m_j}{j!}(ik)^j}=1-{\frac{\sigma^2
k^2}{2}}+{\frac{m_4k^4}{4!}}+... 
\label{eq:powerSerOfFK}
\end{equation}
where the coefficients of this series are given in terms of the moments
$m_j=\langle x^j \rangle$ of $w(x)$.
In the same way, one can expand $\psi(k)$ in a power series in terms of the
cumulants of $w(x)$:
\begin{equation}
\psi(k) = \sum_{j=1}^{\infty}\frac{\kappa_j}{j!} (ik)^j=-{\frac{\sigma^2
k^2}{2}}+{\frac{1}{4!}}(m_4-3\sigma^4)k^4+...
\label{eq:powerSerOfPsiK}
\end{equation}
where the $j$th cumulant, ${\kappa_j}$, is related to the first $j$ moments by
the following relation \cite{blinnikov1998expansions}:
\begin{equation}
\kappa_j = j!\sum_{\{k_\alpha\}} (-1)^{r-1}(r-1)!
\prod_{\alpha=1}^j\frac{1}{k_\alpha !}\left(\frac{m_\alpha}{\alpha
!}\right)^{k_\alpha}.
\label{eq:cumulantsMoments}
\end{equation}
Here, the summation is
over all sets $\{{k_\alpha}\}$ satisfying  $k_1 + 2k_2 +
\ldots + jk_j = j$, and $r=\sum_{\alpha=1}^j k_\alpha$. Hence, $\kappa_1 =0$
(since for a symmetric $w(x)$ the first moment vanishes),
$\kappa_2=m_2$, etc.
In the last three equations all odd terms in the series expansions vanish, since
$w(x)$ is symmetric.

For the normalized sum $S_n$, an equivalent expansion exists:
\begin{equation}
\psi_{S_n}(k) = n\ln \left(\tilde{w}(k/\sqrt{n})\right) =
\sum_{j=1}^{\infty}\frac{\kappa_j}{j!}
\frac{(ik)^j}{n^{j/2-1}}.
\label{eq:psiSK_expansion}
\end{equation}
Substituting $m_1=0$ (all odd $j$ terms vanish), $\kappa_2=\sigma^2$ and
$s=j-2$, we can rewrite ${\tilde{w}_S}(k)$ as:
\begin{equation}
{\tilde{w}_S}(k) = e^{\psi_S(k)} = e^{-\sigma^2 k^2/{2}} \exp \left[
\sum_{s=1}^{\infty}\frac{\kappa_{s+2}}{(s+2)!}{(ik)^{s+2}}n^{-s/2} \right].
\label{eq:fSK_expansion1}
\end{equation}
Expanding the exponent in a power series in $n^{-1/2}$ we get (all odd $\nu$
terms vanish because of the symmetry):
\begin{equation}
{\tilde{w}_S}(k) = e^{-\sigma^2 k^2/{2}} \left[1+
\sum_{\nu=1}^{\infty}P_{2\nu}(ik)n^{-\nu} \right],
\label{eq:fSK_expansion2}
\end{equation}
where:
\begin{equation}
P_\nu(ik)=\sum_{\{k_\alpha\}} \prod_{\alpha=1}^\nu \frac{1}{k_\alpha!}\left[
\frac{\kappa_{\alpha+2}(ik)^{\alpha+2}}{(\alpha+2)!} \right]^{k_\alpha}.
\label{eq:P_nu_coefficients}
\end{equation}
Here the summation over the set $\{k_\alpha\}$ for a given $\nu$ is defined as
above.
For example, $P_2(ik)={{\kappa_4 k^4}/{4!}}$ and ${P_4(ik)={-{\kappa_6
k^6}/{6!}}}$.

Taking the inverse Fourier transform of ${\tilde{w}_S}(k)$ we get:
\begin{align}
w_S(x) &= Z_\sigma(x)\left[ 1 + \sum_{\nu=1}^{\infty}\frac{q_{2\nu}(x)}{n^{\nu}}
\right]\nonumber\\
&=Z_\sigma(x)\left[1+{\frac{\kappa_4}{4!n}}{\frac{1}{\sigma^4}}
H_4({\frac{x}{\sigma}})+{\frac{\kappa_6}{6!n^2}} {\frac{1}{\sigma^6}}
H_6({\frac{x}{\sigma}})+...\right],
\label{eq:wSX1}
\end {align}
where:
\begin{equation}
q_\nu(x) = \sum_{\{k_\alpha\}}\frac{1}{\sigma^{\nu+2r}} H_{\nu+2r}(x/\sigma)
\prod_{\alpha=1}^\nu
\frac{1}{k_\alpha!}\left(\frac{\kappa_{\alpha+2}}{(\alpha+2)!}\right)^{k_\alpha}
,
\label{eq:qNu}
\end{equation}
where $r$ is defined as above, and $H_n(x)$ is the $n$th order 
Hermite polynomial \cite{abramowitz1972handbook}. For 
example $q_2=\kappa_4/4!\sigma^4 H_4(x/\sigma)$ and $q_4=\kappa_6/6!\sigma^6
H_6(x/\sigma)$ in agreement with Eq. (\ref{eq:wSX1}). This result, known as the
classical Edgeworth expansion \cite{edgeworth1906generalised}, 
was first obtained by Petrov as an infinite series
\cite{blinnikov1998expansions, petrov1975sums}.

The Edgeworth expansion is a true asymptotic expansion of $w_S(x)$ only when
all of the moments of $w(x)$ exist. However, for a heavy-tailed $w(x)$ with a finite
variance (so that the CLT holds), higher moments diverge, and this series
expansion cannot reproduce the behavior of $w_{S_n}(x)$.
Yet, one may consider a truncated series ignoring the higher order diverging
terms. 
This ad hoc truncated series may work well in the central part of $w_S(x)$.
However it completely fails to predict the rare events as we shall show later.
\iffalse
This \textit{truncated Edgeworth series} for finite $n$ still gives, as we will
see, a better
approximation to the exact shape of $w_S(x)$ compared to the standard CLT
approach, at least for the specific cases investigated below.
\fi

\section{Generalization of the Edgeworth Expansion}
\label{sec:levi-edge}

\subsection{The Fractional Generalized Series}
\label{subsec:fracGenSer}

The Edgeworth and the truncated Edgeworth expansions deal only with probability 
densities $w(x)$ with finite variance. For a normalizable
symmetric PDF with a diverging variance, where $w(x)\sim A|x|^{-(1+\alpha)}$
for large $x$ and $0<\alpha<2$ with $w(x)=w(-x)$, the generalized CLT
states~\cite{samoradnitsky1994stable} that in the limit $n\to \infty$, the PDF
of the sum
\begin{equation}
S_n=\sum_{i=1}^nx_j/n^{1/\alpha}
\label{eq:LeviVariableSum}
\end{equation}
approaches the symmetric L\'{e}vy
$\alpha$-stable density function, $L_{\alpha,\tilde{A}}(x)$
\cite{gorska2011levy}:
\begin{equation}
\lim_{n\rightarrow\infty}w_S(x) = L_{\alpha,\tilde{A}}(x)
\equiv \frac{1}{\pi}\int_{0}^{\infty}\cos(kx)\exp(-\tilde{A}k^\alpha)dk,
\label{eq:L_alpha_A}
\end{equation}
where
\begin{equation}
\tilde{A}=\frac{A\pi}{\Gamma(\alpha+1)}\sin(\pi \alpha/2).
\label{eq:A_tilde}
\end{equation}
Hence for the family of PDFs approaching the 
L\'{e}vy $\alpha$-stable density function (as $n\rightarrow \infty$), we cannot
use the Edgeworth series expansions, since the latter's asymptotic behavior is Gaussian.
Long-tailed PDFs can be found in many stochastic processes e.g., 
in polymer physics, fluid dynamics, cold atoms, biophysics, optics,
engineering, economics etc. \cite{barkai2000levy, bouchaud1990anomalous,
cottone2009use, barthelemy2008levy, mantegna2000introduction,
mantegna1995scaling}.
Later on we will analyze
the case of cold atoms in an optical lattice in detail.

Our approach to these PDFs uses a series expansion of $\tilde{w}_S(k)$
which asymptotically goes to the Fourier transform of L\'{e}vy $\alpha$-stable
density function.
Given a normalized symmetric $w(x)$ with a diverging variance, one may expand
${\tilde{w}}(k)$ in a generalized Taylor series \cite{cottone2009use}: 
\begin{equation}
\tilde{w}(k) = 1 + \sum_{i=1}^{\infty} a_i|k|^{\alpha_i},
\label{eq:fk_Levi}
\end{equation}
where $\alpha_i>0$ could be either integer or non-integer powers of $|k|$. In
general the sum can also include terms such as $ln(|k|) |k|^{\alpha_i}$.

Using the same scheme as before, now for the $S_n$ given in Eq.
(\ref{eq:LeviVariableSum}) (here $\alpha\equiv\alpha_1=min\{\alpha_i\}$
is the asymptotic L\'{e}vy exponent), we get:
\begin{align}
\tilde{w}_S(k) & = \exp\big[n
\ln\big(\tilde{w}(k/n^{1/\alpha})\big)\big]\nonumber \\ 
& = \exp{(-a|k|^\alpha) } \bigg[1 + \sum_{r=1}^{\infty}
\frac{b_r}{n^{(\gamma_r/\alpha)-1}}|k|^{\gamma_r} \bigg],
\label{eq:fSk_Levi}
\end{align}
where $a\equiv -a_1$ (since $a_1<0$), $b_r$ are coefficients depending on the
explicit form of $w(x)$, and $\gamma_r$ are the powers of $|k|$ when expanding
the exponential.
In principle, the $\gamma_r$, $b_r$ are determined by $\alpha_i$
and $a_i$, which are in turn obtained from Fourier transform of $w(x)$.
An example for this relation will be given later, when we will deal with the
application of these equations to the special case of cold atoms.
{In what follows, terms of the form $\exp(-a|k|^\alpha)|k|^\gamma$ where
$\gamma$ is not even or where $\alpha<2$ will be called \textit{non-analytic
terms} due to their small $k$ non-analytic behavior.}

Scaling out $a$ by substituting $\tilde{k} = a^{1/\alpha}k$ and
$\tilde{x}=a^{-1/\alpha}x$, the inverse 
Fourier transform of the first term of $\tilde{w}_S(\tilde{k})$ gives
\cite{barndorff2001levy, feller2008introduction}:
\begin{equation}
\frac{1}{\pi} \int_0^\infty \cos(\tilde{k} \tilde{x})
\exp(-\tilde{k}^\alpha) d\tilde{k} =
L_\alpha(\tilde{x}),
\label{eq:p_Levi}
\end{equation}
where $L_\alpha(\tilde{x}) \equiv L_{\alpha,1}(\tilde{x})$, and
$\int_{-\infty}^{\infty}L_\alpha(\tilde{x})d\tilde{x}=1$. Thus the first term
gives the L\'evy CLT as expected.

Each additional term in Eq. (\ref{eq:fSk_Levi}), when transformed back to
$\tilde{x}$ space includes an integral of the form:
\begin{align}
T_{\alpha,\gamma}(\tilde{x})=\frac{1}{\pi} & \int_0^\infty \cos(\tilde{k}
\tilde{x})
\exp(-\tilde{k}^\alpha)
\tilde{k}^\gamma d\tilde{k}.
\label{eq:T_ag1}
\end{align}
These expressions where introduced also in the context of  L\'evy
Ornstein-Uhlenbeck process  
\cite{Janakiraman2014unusual,jespersen1999levy,toenjes2013nonspectral}.

A term of this form can be written as a derivative of order $\gamma$ (not
necessarily integer) of $L_\alpha$ and of what we call the
\textit{conjugate} L\'{e}vy function:
\begin{equation}
R_{\alpha}(\tilde{x}) = \frac{1}{\pi}\int_0^{\infty} \sin(\tilde{k} \tilde
{x}) \exp(-\tilde{k}^\alpha) d\tilde{k},
\label{eq:p_conjLevi}
\end{equation}
such that:
\begin{equation}
T_{\alpha,\gamma}(\tilde{x}) = \nu_1\left(_{\tilde{x}} D^{\gamma}_{\infty}
\left[L_\alpha(\tilde{x}) \right] \right) - \nu_2\left(_{\tilde{x}}
D^{\gamma}_{\infty}
\left[R_\alpha(\tilde{x}) \right] \right),
\label{eq:rimanLewvilleOperator}
\end{equation}
where $\nu_1 = \cos(\frac{\gamma \pi}{2})$ and $\nu_2 = \sin(\frac{\gamma
\pi}{2})$. In Eq. (\ref{eq:rimanLewvilleOperator}) we have used the
Weyl-Reimann-Liouville
\cite{lovoie1976fractional, mathai2009h} definition for the fractional
derivative $_{\tilde{x}}
D^{\gamma}_{\infty}$ (for more details, see {Appendix
\ref{app::Fractional})}.
This expression holds both for integer (odd and even) and for non-integer
$\gamma$. When $\gamma$ is an even integer (in which case we replace $\gamma$
with $2n$), the second term vanishes and the first term
reduces to $(-1)^{n}d^{2n}/d{\tilde{x}}^{2n} L_\alpha({\tilde{x}})$.
For odd integer $\gamma=2n+1$,
on the other hand, the first term vanishes and the second term reduces to
$(-1)^{n}d^{2n+1}/d{\tilde{x}}^{2n+1} R_\alpha(\tilde{x})$.

The inverse Fourier transform of Eq. (\ref{eq:fSk_Levi}) written
in terms of $T_{\alpha,\gamma}({\tilde{x}})$ gives:
\begin{equation}
w_S(\tilde{x}) = \left[
L_\alpha(\tilde{x})+\sum_{r=1}^{\infty}\frac{b_r}{a^{\gamma_r/\alpha}n^{
\gamma_r/\alpha-1}}T_{\alpha,\gamma_r}(\tilde{x}) \right].
\label{eq:wSx_Levi}
\end{equation}
Indeed, this expansion is general in the sense that it covers both the L\'{e}vy
regime (where the variance diverges) and the Gaussian regime.
This expansion in the Gaussian regime includes two cases: (i) the
case where all moments exist (the classical Edgeworth expansion); and (ii)  the
case where only a finite number of higher moments exist (i.e., the
\textit{fractional Gauss
Edgeworth expansion}). In the L\'{e}vy regime, $\alpha<2$, we call the
expression in Eq. (\ref{eq:wSx_Levi}) the \textit{fractional L\'evy Edgeworth
expansion}.
In the Gaussian regime, $\alpha=2$, one gets
$L_{2}(\tilde{x})=Z_{\sigma}(\tilde{x})$ (where $\tilde{x}=x/a^{1/\alpha}$
is related to $\sigma$ by: $a=\sigma^2/2$) and
$I_2(\tilde{x})=(1/\pi)Daw(\tilde{x}/2)$ ($\textrm{$Daw$}(\cdot)$ is the Dawson
function \cite{abramowitz1972handbook}), and the following
$T_{2,\gamma_r}(\tilde{x})$ terms could be either regular integer-order or
fractional derivatives of $L_2(\tilde{x})$ and $I_2(\tilde{x})$ \footnote{The
fractional derivatives of $L_2(\tilde{x})$ and $I_2(\tilde{x})$ can be written
in terms of the parabolic-cylindric functions (see
e.g., \cite{abramowitz1972handbook}) as presented in Eq. (\ref{eq:parab})}. As a
result, when not all the moments exist, the density function in Eq.
(\ref{eq:wSx_Levi}) reduces to the fractional Gauss Edgeworth expansion. In the
Gaussian regime, there exists an exception, i.e., for PDFs  of the form
$w({x}) \sim {x}^{-(1+\alpha)}$ for large ${x}$, when $\alpha$
is an even integer. In this particular case, $\tilde w(k)$ contains terms such
as $\exp(-a|k|^\alpha)|k|^\gamma\ln|k|$, and one has to define the special
$T^{ln}_{\alpha,\gamma}$ function:
\begin{equation}
 T^{ln}_{\alpha,\gamma}(\tilde{x})=\frac{1}{\pi} \int_0^\infty \cos(\tilde{k}
 \tilde{x})
 \exp(-\tilde{k}^\alpha)\ln(\tilde{k}) \tilde{k}^\gamma d\tilde{k}.
 \label{eq:T_ag3}
\end{equation}
 For the analysis of this particular case, see, e.g.,
Ref. \cite{lam2011corrections}.

\subsection{Further Investigation of $T_{\alpha,\gamma}$}
\label{subsec:TalphaGamma}

In order to reveal the behavior of these series in the limits of large and
small $\tilde{x}$, it is instructive to express $T_{\alpha,\gamma}({\tilde{x}})$
in terms of
$H$-Fox functions \cite{mathai2009h}. Moreover, since
$T_{\alpha,\gamma}({\tilde{x}})$ is the sum of fractional derivatives of
$L_{\alpha}(\tilde{x})$ and $R_{\alpha}(\tilde{x})$, it is convenient to
express them in terms of  
$H$-Fox functions, because a fractional derivative of an $H$-function is another
$H$-function with shifted indices \cite{mathai2009h}. We discuss these functions
in detail in Appendix \ref{app::hFox}, and show their relation to the fractional
derivatives in Appendix \ref{app::Fractional}.
Since $T_{\alpha,\gamma}(\tilde{x}) = Re[\int_0^\infty \exp(-i\tilde{k}
\tilde{x})
\exp(-\tilde{k}^\alpha) \tilde{k}^{\gamma} d\tilde{k}]$, and using the
Mellin transform:
$\exp(-i\tilde{k}\tilde{x})=1/(2\pi i)\int_L \Gamma(s) (i\tilde{k}
\tilde{x})^{-s} ds$ and integrating over
$\tilde{k}$ (for more details, see Appendix \ref{appsub:tAlphaGamma}), one may
express $T_{\alpha,\gamma}(\tilde{x})$ as the Mellin-Barnes integral
\cite{braaksma1936asymptotic}:
\begin{equation}
T_{\alpha,\gamma}(\tilde{x})= \frac{1}{2\pi i \alpha} \int_L
\frac{\Gamma(s)\Gamma(\frac{1+\gamma-s}{\alpha})}{\Gamma(\frac{1-s}{2}
)\Gamma(\frac{1+ s}{2})}x^{-s}ds.
\label{eq:T_ag4}
\end{equation}
By definition, this integral is an $H$-Fox function
\cite{mathai2009h, mainardi2005fox}:\\
\begin{equation}
T_{\alpha,\gamma}(\tilde{x}) = \left\{\begin{array}{ll}\rule[5pt]{0mm}{15pt}\frac{1}{\alpha}H^{1,1}_{2,2}
\left[ \frac{1}{\tilde{x}} \bigg| \begin{array}{cc}\rule[2pt]{0mm}{11pt}
{(1,1),(\frac{1}{2}, \frac{1}{2})} \\\rule[2pt]{0mm}{11pt}
{(\frac{1+\gamma}{\alpha},\frac{1}{\alpha}),(\frac{1}{2},\frac{1}{2})}
\end{array}
\right] ; & 0<\alpha<1 \\ %\\ \vspace*{-0.2in}
\rule[5pt]{0mm}{15pt} \frac{1}{\alpha}H^{1,1}_{2,2}
\left[ \tilde{x} \bigg| \begin{array}{cc}\rule[2pt]{0mm}{11pt}
{(1-\frac{1+\gamma}{\alpha},\frac{1}{\alpha}),(\frac{1}{2},
\frac{1}{2})} \\ \rule[2pt]{0mm}{11pt}
{(0,1),(\frac{1}{2},\frac{1}{2})} \end{array}
\right] ; \qquad& 1<\alpha\leq 2\end{array}\right.
\label{eq:T_ag5_hFox2_1}
\end{equation}
For $\gamma=0$, Eq. (\ref{eq:T_ag5_hFox2_1}) 
reduces to the $H$-Fox function representing the
symmetric L\'{e}vy $\alpha$-stable density function, and also for
$\gamma=0$ and $\alpha=2$ to the Gaussian density function
\cite{schneider1986stable,mainardi2005fox}.

For $\alpha=2$ and even $\gamma$, one returns to the appropriate
Gauss-Hermite function (see e.g., Fig. \ref{fig:Fig3})
\begin{align}
T_{2,\gamma}(\tilde{x}) 
& = \frac{1}{2}H^{1,1}_{2,2}
\left[ \tilde{x} \bigg| \begin{array}{cc}
{(1-\frac{1+\gamma}{2},\frac{1}{2}),(\frac{1}{2},
\frac{1}{2})} \\ {(0,1),(\frac{1}{2},\frac{1}{2})} \end{array}
\right] \nonumber\\
&\qquad=
\frac{1}{2^{\gamma/2}} \frac{1}{2\sqrt{\pi}} e^{-\frac{{\tilde{x}^2}}{4}}
H_\gamma({\frac{\tilde{x}} {\sqrt{2}}}), \nonumber\\
&\qquad=
\frac{1}{2^{\gamma/2}} Z_{\sqrt{2}}(\tilde{x})
H_\gamma({\frac{\tilde{x}} {\sqrt{2}}}),
\label{T_ag5_hFox2_2}
\end{align}
which is the regular term generated by the Edgeworth expansion in Eq.
(\ref{eq:wSX1}).

One may extract the behavior of $T_{\alpha,\gamma}(\tilde{x})$ for large and
small $\tilde{x}$ values.
In Appendix \ref{app::hFox} we derive the following series for
$T_{\alpha,\gamma}(\tilde{x})$.
\iffalse
Closing the contour in Eq. (\ref{eq:T_ag4}) on the
left hand side by a semi-circle of radius $R \rightarrow \infty$, we obtain a
asymptotic 
series expansion of $H(\tilde{x})$ in (positive) powers of $\tilde{x}$ which
describe the small
$\tilde{x}$ regime, while closing the contour on its right hand side, we get an
asymptotic expansion
in inverse powers of $\tilde{x}$ which describes the large $\tilde{x}$ regime
\cite{garoni2002levy}.
\fi
For the small $\tilde{x}$ regime, when $\alpha>1$ we get:
\begin{align}
T_{\alpha,\gamma}(\tilde{x}) =
\frac{1}{\alpha\pi}\sum_{n=0}^{\infty}\frac{(-1)^n}{ \Gamma(1+{2n}) }
\Gamma\left(\frac{1+\gamma+{2n}}{\alpha}\right)\tilde{x}^{2n} ,
\label{eq:T_ag6}
\end{align}
while for large $\tilde{x}$ when $0<\alpha<1$ we get:
%\begin{align}
\begin{equation}
T_{\alpha,\gamma}(\tilde{x}) =
 \frac{1}{\pi}\sum_{n=0}^{\infty} 
\bigg[\frac{(-1)^n\Gamma\left({1+\gamma+n\alpha}\right)}{\Gamma(1+n)}
%\nonumber\\ & \qquad\times
\cos\left(\frac{
1+\gamma+n\alpha } { 2 }\pi \right)\tilde{x}^{-(1+\gamma+n\alpha)} \bigg] ,
\label{eq:T_ag7}
\end{equation}
%\end{align}
Using Mathematica for selected values of rational pairs of $\alpha$
and $\gamma$ this series could be represented by combinations of special
functions.
Each of these series expansions for $T_{\alpha,\gamma}$ is a converging series
for the suitable $\alpha$ range as mentioned above. However, one may still use
the small $\tilde{x}$ expansion for $0<\alpha<1$ and vice versa, while
remembering that in that range of $\alpha$s the series is an asymptotic
one.

\begin{figure}[tb]
\center{\includegraphics[width=0.7\textwidth]{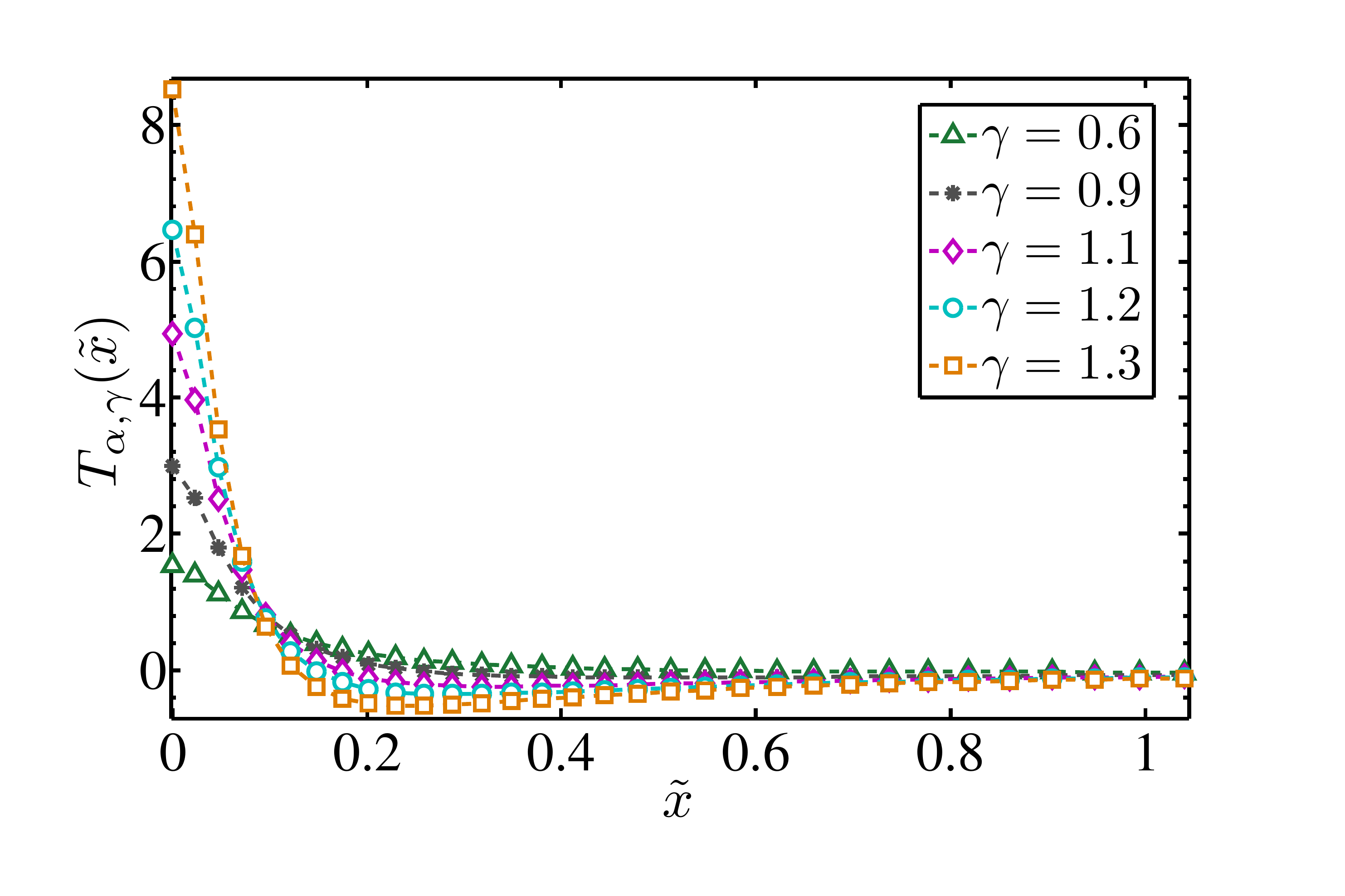}}
\caption{$T_{\alpha,\gamma}(\tilde{x})$ for $\alpha=0.5$,
and for various $\gamma$ values. The $T_{\alpha,\gamma}(\tilde{x})$ values were
calculated both by numerical inverse Fourier transform of the expression in Eq.
(\ref{eq:T_ag1}) (markers) and by calculating the series expansion in Eq.
(\ref{eq:T_ag7}) with $2\cdot10^5$ terms (dashed lines). Note that we only
illustrate the regime $\gamma>\alpha$, since these are the terms that appear in
our series expansion.}
\label{fig:Fig1}
\end{figure}

\begin{figure}[tb]
\center{\includegraphics[width=0.7\textwidth]{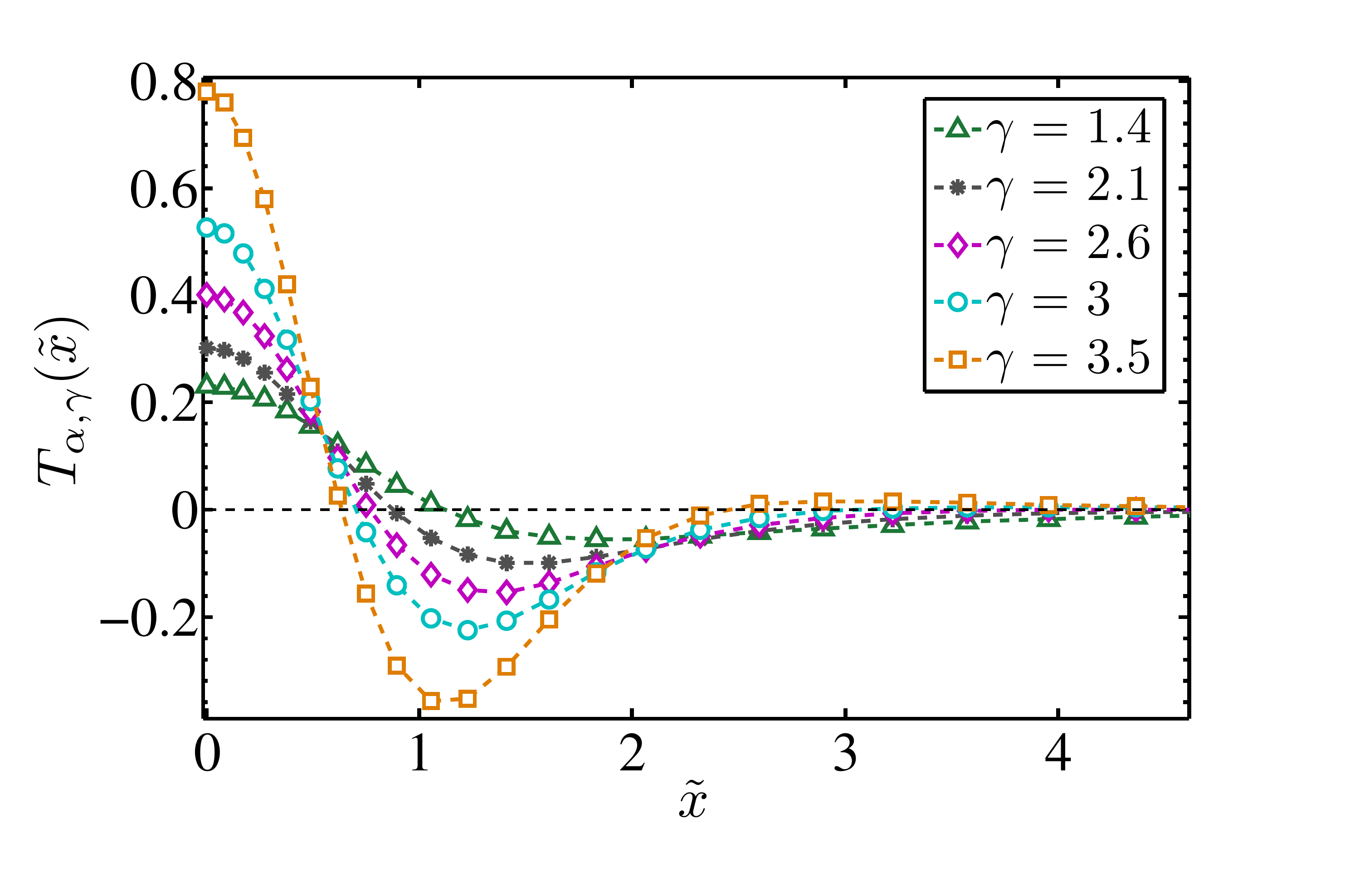}
}
\caption{$T_{\alpha,\gamma}(\tilde{x})$ for $\alpha=1.3$,
and for various $\gamma$ values. The $T_{\alpha,\gamma}(\tilde{x})$ values were
calculated as in Fig. \ref{fig:Fig1}.}
\label{fig:Fig2}
\end{figure}

\begin{figure}[tb]
\center{\includegraphics[width=0.7\textwidth]{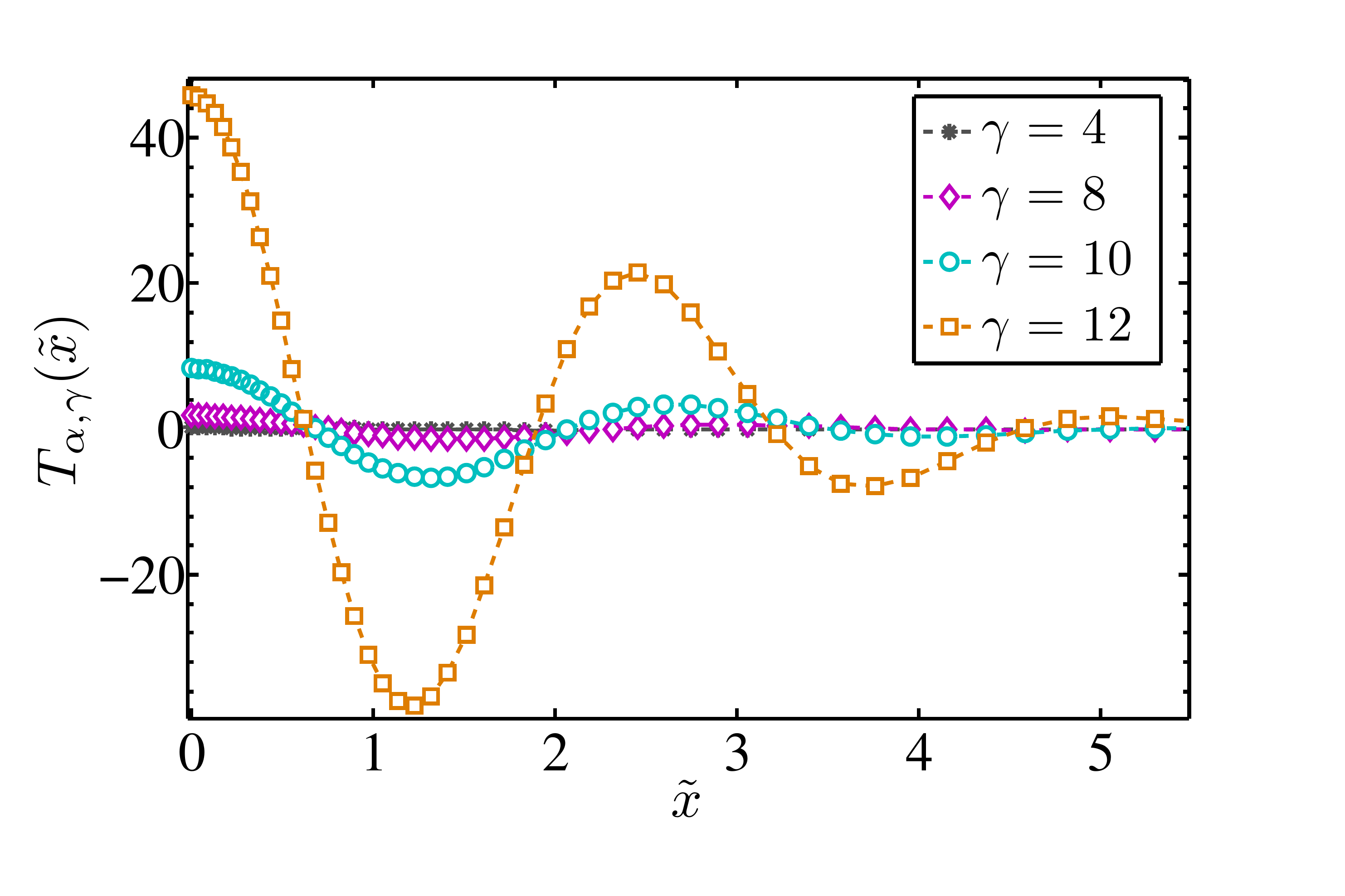}
}
\caption{$T_{\alpha,\gamma}(\tilde{x})$ for $\alpha=2$ and even
$\gamma$. These functions corresponds to the Gauss-Hermite functions. 
The $T_{\alpha,\gamma}(\tilde{x})$ values were
calculated as in Fig. \ref{fig:Fig1}.}
\label{fig:Fig3}
\end{figure}

In order to better understand the dependence of $T_{\alpha,\gamma}(\tilde{x})$
on $\alpha$ and $\gamma$, we plot $T_{\alpha,\gamma}(\tilde{x})$ for different
values of $\alpha$ and $\gamma$. In Fig. \ref{fig:Fig1}
we plot $T_{\alpha,\gamma}(\tilde{x})$ for $\alpha=0.5$ and different
$\gamma$. In Fig. \ref{fig:Fig2}
we plot $T_{\alpha,\gamma}(\tilde{x})$ for $\alpha=1.3$ and different
$\gamma$. As one can observe, these terms are positive at the center, and
decreasing until they become negative, and then increasing again so that
asymptotically they tend to zero.

$\gamma$ affects the amplitude of $T_{\alpha,\gamma}(\tilde{x})$, i.e., the
maximal (at $\tilde{x}=0$) and minimal (negative) values of
$T_{\alpha,\gamma}(\tilde{x})$. One
has to keep in mind that for $\gamma\rightarrow 0$ we get
$T_{\alpha,\gamma}(\tilde{x})\rightarrow L_\alpha(\tilde{x})$, which is always
positive. This implies that as $\gamma$ decreases the negative area also
decreases, and
since $\int_0^\infty T_{\alpha,\gamma}(\tilde{x})d\tilde{x}=0$ (for all
$\gamma$ except for $\gamma=0$) since  $\gamma=0$ yields a ``pure''
L\'evy $\alpha$-stable density function which is normalized and the
additional terms in the series must preserve the normalization, so that the
integral
over them must vanish. As can be seen in Fig. \ref{fig:Fig1} and Fig.
\ref{fig:Fig2}, when $\gamma$ decreases, both the positive and negative parts
of $T_{\alpha,\gamma}(\tilde{x})$ decrease, and in addition the value of
$\tilde{x}$ where $T_{\alpha,\gamma}(\tilde{x})$ crosses the $\tilde{x}$-axis
increases (so that for $\gamma \rightarrow 0$ this value should go to infinity,
to recover the positive definite $L_\alpha(\tilde{x})$ PDF). 

\begin{figure}[tb]
\center{\includegraphics[width=0.7\textwidth]
{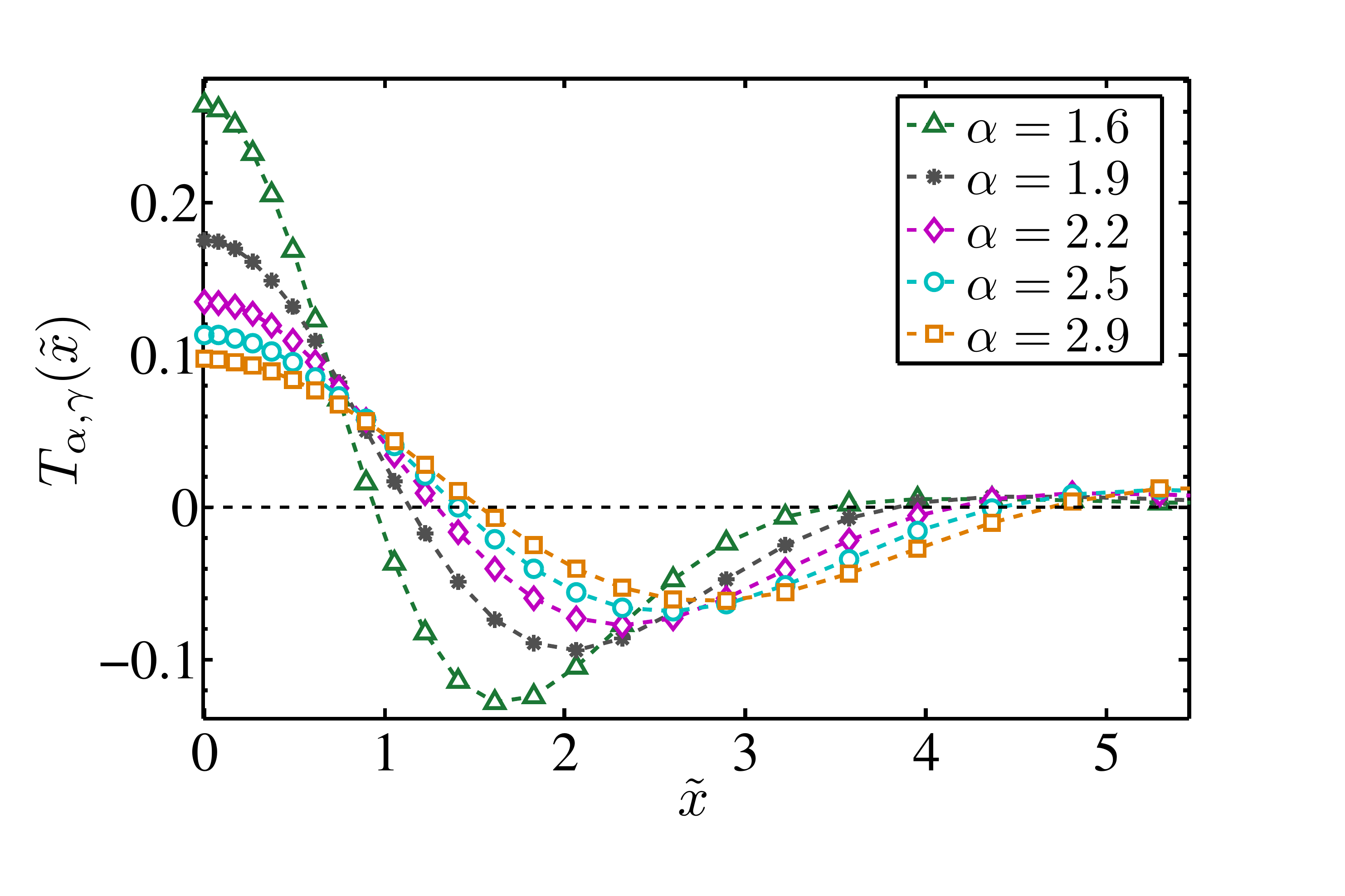}}
\caption{$T_{\alpha,\gamma}(\tilde{x})$ for $\gamma=3$ and various $\alpha$
values. The $T_{\alpha,\gamma}(\tilde{x})$ values were
calculated as in Fig. \ref{fig:Fig1}.}
\label{fig:Fig4}
\end{figure}

When trying to characterize the effect of increasing $\alpha$ on
$T_{\alpha,\gamma}$, we first refer to the well studied special case
of $\gamma=0$. In this case, for $\alpha<2$ we get $T_{\alpha,\gamma}(\tilde{x})
= L_\alpha(\tilde{x})$. Increasing $\alpha$ lowers the central peak and widens
the central region of the density function until for $\alpha=2$ this term
becomes a Gaussian.
The same behavior also holds for $\gamma>0$ where increasing $\alpha$ makes the
central peak lower, but the central region becomes wider. The value of
$\tilde{x}$ where $T_{\alpha,\gamma}(\tilde{x})$ crosses the $\tilde{x}$-axis
(which for the positive definite L\'evy and Gaussian is going to infinity)
increases as can be seen in Fig. \ref{fig:Fig4}.

\iffalse
(where for $\alpha=2$ one
gets the Gauss-Hermite functions described above) as can be seen in Fig.
\ref{fig:Fig3}. In addition,
\fi

\subsection{Leading Order Fractional Edgeworth Expansion} 
\label{sec:hazut-correction}
Our approach for analyzing $w_S(x)$ both in the Gaussian and L\'{e}vy regimes is
based on the fact that higher terms in the series expansion in Eq.
(\ref{eq:wSx_Levi}) decrease more
rapidly with $n$. 
We first consider a case where the
Gaussian CLT applies, however, ${w}(x)\sim x^{-(\alpha+1)}$,
where $\alpha>2$, so that sufficiently high order moments diverge. The truncated
Edgeworth
expansion will not give a good estimate of the tail of the PDF. The tail is
described by non-analytical terms in the $k$ expansion, while the Gaussian CLT
and the truncated Edgeworth expansion rely on analytical terms of the type
$k^{2n}\exp(-k^2)$ when $n$ is an integer.
For example, the function $w(x)=3/2{\pi}(1+x^6)$ is expanded in the Fourier
space to:
\begin{equation}
\tilde{w}(k) = 1 - \frac{k^2}{4} + \frac{k^4}{24} - \frac{|k|^5}{80} + ...
\label{exampleFourier}
\end{equation}
where the three terms (up to the $k^4$ term) are analytic, and the terms
from the $|k|^5$ belong to the non-analytical part.

In the leading order fractional Edgeworth expansion we neglect all terms in
the series that are higher than the first non-analytic term. In the L\'evy
regime, all the terms are non-analytic ($\alpha<2$), since even the second
moment diverges
and in this regime we take only the first term of the series. In the Gaussian
regime (finite variance,
$\alpha=2$), however, we need to take all the analytic terms first (the
truncated Edgeworth series), but since these terms do not capture the behavior
of the (heavy-) tailed nature of the $w(x)$ (its diverging moments) we still
need to add the first non-analytic term in order to capture the power-law
decay of the tails.

In the latter case (Gaussian regime), for PDFs that decay as
$Ax^{-(1+\alpha)}$ ($\alpha>2$) for large $x$, this approach yields:
\begin{equation}
{\tilde{w}_S}(k) = e^{-\sigma^2 k^2/{2}} \left[1+
\sum_{\nu=1}^{\nu<\alpha}P_{2\nu}(ik)n^{-\nu} + \frac{1}{n^{\alpha/2-1}}\xi(k)
\right],
\label{eq:fSK_expansion3}
\end{equation}
where $P_\nu$ is defined in Eq. (\ref{eq:P_nu_coefficients}) and the summation
of
the
Edgeworth part (second term in the brackets) is only
over even values of $\nu$ (because of the symmetric nature of $w(x)$) and is
over values of $\nu$
up to but not including $\alpha$. The last term
is given by:
\begin{equation}
\xi(k) = 
\left\{\begin{array}{ll}\rule[5pt]{0mm}{10pt}
-\frac{A\pi}{\Gamma(\alpha+1)\sin(\alpha\pi/2)}|k|^\alpha, & \alpha\neq 2n 
\\ \rule[5pt]{0mm}{15pt} 
\frac{2A\pi(-1)^{\alpha/2}}{\Gamma(\alpha+1)}|k|^{\alpha}\log|k|,
& \alpha=2n 
\end{array}\right.
\label{eq:xi}
\end{equation}

The corresponding $w_S(x)$
is then:
\begin{equation}
w_S(x) = Z_\sigma(x) +
\sum_{\nu=1}^{\nu<\alpha}\frac{\zeta_{2\nu}\left(\frac{\sqrt{2}x}{\sigma}\right)
}{n^{\nu}} +
\frac{2^{(1+\alpha)/2}b_\alpha}{n^{\alpha/2-1}\sigma^{1+\alpha}}
T_{2,\alpha}\left(\frac{\sqrt{2}x}{\sigma}\right).
\end{equation}
where $\zeta_{\nu}$ is defined using the $T_{\alpha,\gamma}$ terms:
%\begin{align}
\begin{equation}
\zeta_\nu\left(\frac{\sqrt{2}x}{\sigma}\right) %& 
=
\left(\frac{\sqrt{2}}{\sigma}\right)^{\nu+2r+1}
\sum_{\{k_m\}}T_{2,\nu+2r}\left( \frac{\sqrt{2}x}{\sigma} \right)% \nonumber\\
%& \qquad\qquad\qquad\times
\prod_{m=1}^\nu
\frac{1}{k_m!}\left(\frac{\kappa_{m+2}}{(m+2)!}\right)^{k_m},
\label{eq:fSK_expansion}
\end{equation}
%\end{align}
where $r$ and the sets $\{k_m\}$ are the same as in Eq.
(\ref{eq:cumulantsMoments}), from Eq. (\ref{eq:xi}):
\begin{equation}
b_\alpha =
\left\{\begin{array}{ll}\rule[5pt]{0mm}{10pt}
-\frac{A\pi}{\Gamma(\alpha+1)\sin(\alpha\pi/2)}, & \alpha\neq 2n 
\\ 
\rule[5pt]{0mm}{15pt}
\frac{2A\pi(-1)^{\alpha/2}}{\Gamma(\alpha+1)},
&  \alpha=2n
\end{array}\right.
\label{eq:b_alpha}
\end{equation}
where for the case of even $\alpha$, the $T_{2, \alpha}$ is defined as in Eq.
(\ref{T_ag5_hFox2_2}).

\begin{figure}[tb]
\center{\includegraphics[width=0.7\textwidth]{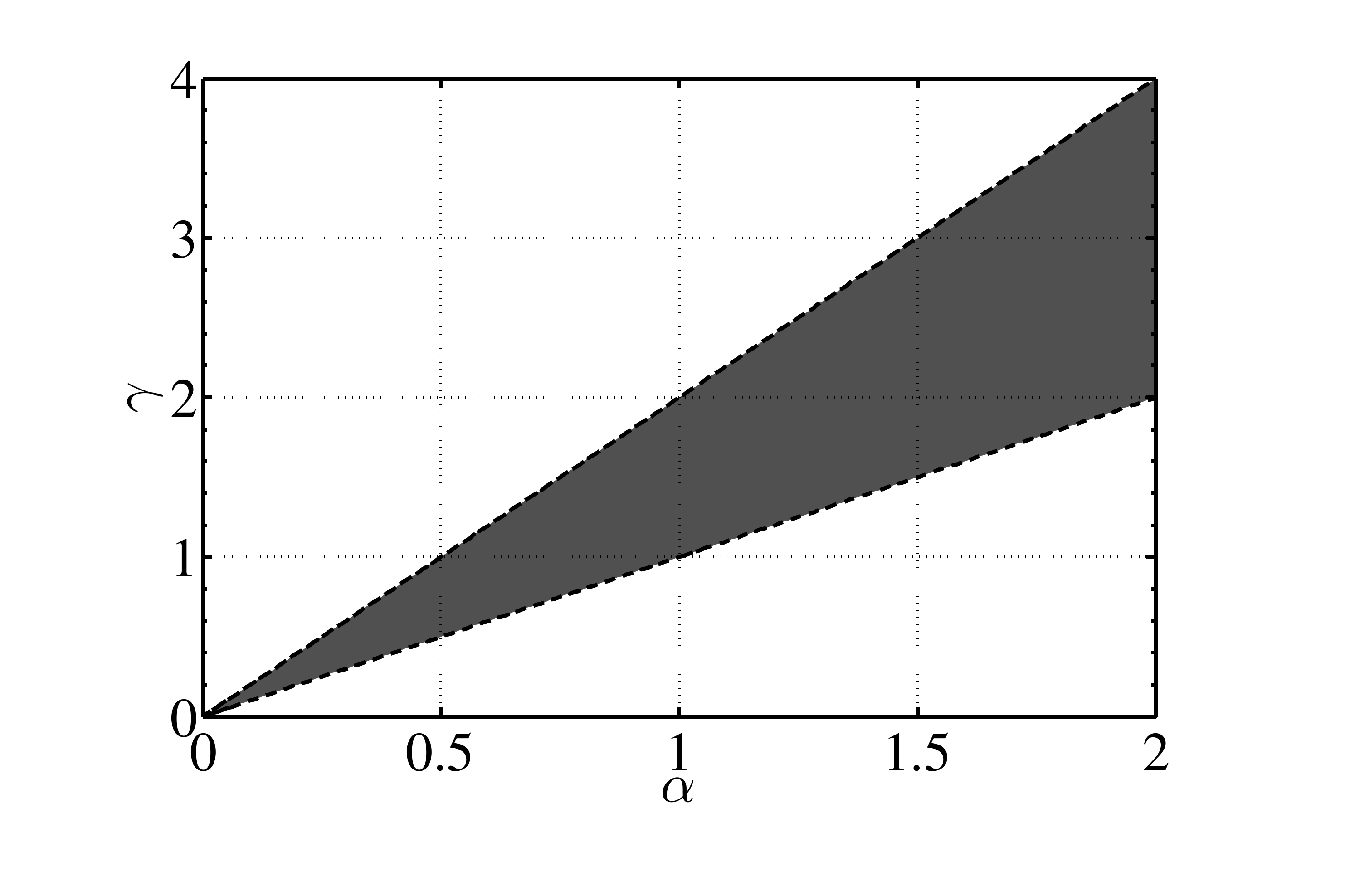}}
\caption{The strip of possible $\gamma$ values for the first
correction term, $T_{\alpha,\gamma}(\tilde{x})$
as a function of $\alpha$ shown as the shaded area in the figure.}
\label{fig:gammaPhase}
\end{figure}

In the L\'{e}vy regime ($0<\alpha<2$), as mentioned, all the cumulants diverge.
As a consequence there are no terms in the Edgeworth expansion, and only
the non-analytic terms exists.
We used the same scheme as in Eq. (\ref{eq:fk_Levi}) to Eq. (\ref{eq:wSx_Levi}).
In Eq. (\ref{eq:fSk_Levi}) we expand $\tilde{w}_S(k)$ in a power
series in two stages: first, we use the expansion $\tilde{w}(k)=
1 + a_1|k|^{\alpha_1} + a_2|k|^{\alpha_2} + O(|k|^{\alpha_2})$, where $\alpha
\equiv \alpha_1$, $a\equiv -a_1$ is a positive constant depending only on
$\alpha$, $a = \pi/[\Gamma(\alpha+1)\sin(\alpha \pi/2)]$
\cite{klafter2011first}, $\alpha_2>\alpha$, and $a_2$ is a constant  depending
on the explicit form of $w(x)$. Then we expand the $\ln(1+x)\simeq x - x^2/2$
and truncate the series after its second term: 
\begin{equation}
\tilde{w}_S(k)\simeq e^{-a|k|^\alpha} \left[ 1+C_n|k|^\gamma \right]
\label{eq:fSk_correctionLevi}
\end{equation}
where:
\begin{equation}
\gamma = \begin{cases}
\alpha_2, & \alpha_2<2\alpha  \\
2\alpha, &  \alpha_2\geq 2\alpha \\
\end{cases},
\label{eq:gamma_correctionLevi}
\end{equation}
and:
\begin{equation}
C_n = \frac{1}{n^{\gamma/\alpha-1}}
\begin{cases}
a_2, &  \alpha_2<2\alpha  \\
-{a^2}/{2}, &  \alpha_2>2\alpha \\
a_2-{a^2}/{2}, &  \alpha_2=2\alpha
\end{cases},
\label{eq:CnCorrectionLevi}
\end{equation}
and the leading order fractional L\'evy Edgeworth takes the
form:
\begin{equation}
w_S(x)\simeq \frac{1}{a^{1/\alpha}}\left[L_{\alpha}(a^{-1/\alpha}x)+
\frac{C_n}{a^{\gamma/\alpha}} T_{\alpha,\gamma}(a^{-1/\alpha}x) \right],
\label{eq:wSx_correctionLevi}
\end{equation}
where $\gamma$ and $C_n$ depend on the explicit form of $w(x)$ as in Eq.
(\ref{eq:gamma_correctionLevi}) and Eq. (\ref{eq:CnCorrectionLevi}).
The possible values of $\gamma$ for a given $\alpha$ value taken from Eq.
(\ref{eq:gamma_correctionLevi}) are shown
in Fig. \ref{fig:gammaPhase}. As discussed in the previous section,
$T_{\alpha,\gamma}(x)$ is positive in its central part, and negative in the
edges. The effect of the leading order fractional L\'evy Edgeworth term on the
PDF depends on the coefficient $C_n$ in Eq. (\ref{eq:wSx_correctionLevi}). If
$C_n$ is positive, this correction will increase the probability for the small
$x$ values, and decrease the probability of large values. If $C_n$ is negative,
the effect will be opposite.

\section{Example: Cold Atoms in Optical Lattice}
\label{sec::ColdAtoms} 
An important physical application of the above methods is in the field of
atoms in an optical lattice undergoing diffusion in momentum space.
\cite{castin1991light, marksteiner1996anomalous}. It has been shown that the
atoms are subjected to a cooling
force (in dimensionless units \cite{kessler2010infinite}) of the form:
\begin{equation}
F(p) = -\frac{p}{1+p^2},
\label{eq:fp}
\end{equation}
where $p$ is the dimensionless momentum of the atom. This cooling force acts to
decrease the momentum of the atom to zero while the fluctuations in momentum can
be
treated as a diffusive process (in momentum space) causing heating. In the
semi-classical picture, one may describe the PDF of the momentum of an atom as
the solution of a Fokker-Planck equation (given, e.g., in
\cite{kessler2010infinite}). The equilibrium solution of this equation,
$W_{eq}(p)$, is given by:
\begin{equation}
W_{eq}(p) = \mathcal{N} (1+p^2)^{-1/(2D)}.
\label{eq:WeqP}
\end{equation}
Here, $\mathcal{N} = \Gamma(\frac{1}{2D}) /[\sqrt{\pi} \Gamma(\frac{1-D}{2D})]$
is a normalization constant, and $D$, the dimensionless
diffusion constant, is defined by $D=cE_R/U_0$, where $U_0$ is the depth of the
optical potential, $E_R$ the recoil energy depends on the
atomic transition involved \cite{cohen1990new,castin1991light}.
Laser cooling experiments indeed show this kind of steady state
solution, where $D$ can be tuned during the experiment to achieve different
steady state behavior~\cite{douglas2006tunable}. The transition between normal
(Gaussian) and anomalous (L\'{e}vy) diffusion in space is also observed
\cite{katori1997anomalous}.

In what follows, we derive the approximate density function for the sum
of the momenta of $n$ such atoms scaled by the appropriate $n^{1/\alpha}$ where
$\alpha$ depends on the value of $D$, as will be explained later. For different
values of $D$ there are three different types of $W_{eq}(p)$. For $D>1$ this
function is not normalizable, and we will not analyze this case further. For
$D<1$, however, there are still two possibilities, the Gaussian regime
$0<D<1/3$ where the variance is finite, $\sigma^2 = D/(1-3D)$, and the
L\'{e}vy regime
$1/3<D<1$, where the variance diverges. The characteristic function of
$W_{eq}(p)$ is:
\begin{equation}
\tilde{w}(k) = \frac{2^{3/2-1/2D}}{\Gamma(1/2D-1/2)}
|k|^{\frac{1}{2D}-\frac{1}{2}}
K_{\frac{1}{2D}-\frac{1}{2}}(|k|),
\label{eq:fkAtoms1}
\end{equation}
where $K$ is the modified Bessel function of the second kind, defined
as:
\begin{equation}
K_\nu(k)=\frac{\pi}{2}\frac{I_{-\nu}(k)-I_{\nu}(k)}{ \sin(\nu \pi) },
\label{eq:fkAtoms2}
\end{equation}
and $I_{\nu}(k)$ is the modified Bessel function of the first
kind. with the Froebenius expansion:
\begin{equation}
I_\nu(k) = 
\sum_{m=0}^{\infty}\frac{1}{m!\Gamma(m+\nu+1)}\left(\frac{k}{2}\right)^{2m+\nu}.
\label{eq:41}
\end{equation}
This series expansion is valid only for non-integer values of $\nu$. The
integral $\nu$ case can be treated as the limit of the non-integral one using
the methods in \cite{abramowitz1972handbook}. Integer values of $\nu$ appear
when $D=1/(2n+1)$ ($n$ is a positive integer) i.e, in the Gaussian regime. For
these
specific $D$ values the series expansion of the modified Bessel $K$ contains
logarithmic terms. For example, for $D=1/5$ we get:
\begin{equation}
\tilde{w}(k)=\frac{1}{4}k^2K_2(|k|)=1-\frac{k^2}{4}+\frac{1}{64}
\bigg(3-4\gamma_E+4\log(2) -4\log(|k|) \bigg)k^4 + ...,
\label{eq:exampleLogarithmic}
\end{equation}
where $\gamma_E\simeq 0.5772$ is the Euler–Mascheroni constant. As mentioned
above, these cases will not be treated here.
 
For this kind of power-law decaying PDF, even in the Gaussian regime that will
be presented below, the Edgeworth series does not converge, since higher
moments do not exist.
Using Eq. (\ref{eq:41}) and defining $\nu=1/2D-1/2$ we get:  
% \begin{equation}
% \tilde{w}(k) = 1-\frac{\Gamma(1-\nu)}{\Gamma(1+\nu)2^{2\nu}}|k|^{2\nu}+
% \frac{\Gamma(1-\nu)}{\Gamma(2-\nu)2^{2}}|k|^2 + ...,
% \end{equation}
\begin{align}
\tilde{w}(k) & =
 \sum_{m=0}^{\infty}\bigg[\frac{\Gamma(1-\nu)}{m!\Gamma(m-\nu+1)}
\left(\frac{|k|}{2}\right)^{2m} \nonumber  -
\frac{\Gamma(1-\nu)}{m!\Gamma(m+\nu+1)}
\left(\frac{|k|}{2}\right)^{2(m+\nu)} \bigg] \nonumber\\
& = 1-\frac{\Gamma(1-\nu)}{\Gamma(1+\nu)2^{2\nu}}|k|^{2\nu}+
\frac{\Gamma(1-\nu)}{\Gamma(2-\nu)2^{2}}|k|^2 + \ldots .
\end{align}

To analyze this further, we need to break it down to two cases. When $2\nu>2$
(which occurs when $0<D<1/3$) we are in the Gaussian regime, with the  leading
order term $\exp(-\sigma^2k^2/2)$ where $\sigma^2=D/(1-3D)$.
\begin{equation}
%\begin{align}
{\tilde{w}_S}(k) \simeq %& 
e^{-\sigma^2k^2/2}\bigg[1+ \sum_{n=1}^{n<2\nu}P_{2n}(ik) 
%\nonumber\\& \qquad\qquad
-\frac{\Gamma(1-\nu)}{\Gamma(1+\nu)2^{2\nu}n^{\nu-1}}|k|^{2\nu} \bigg],
\label{eq:wSk_completeSeries}
\end{equation}
%\end{align}
where $P_{2n}(ik)$ is defined by Eq. (\ref{eq:P_nu_coefficients}) and the sum
runs over all the even powers of $k$ from $4$ to the maximal even integer
smaller than $2\nu$. This truncated Edgeworth correction will vanish (so that
there are no
analytic terms) at the point where the $4$th moment of $w(x)$ diverges,
(i.e., for $D>1/5$). 
It is easy to show that the last term in this equation (the correction term) is
a special case of Eq. (\ref{eq:xi}).
%since from Eq. (\ref{\label{eq:WeqP}), $W_{eq}(p) \sim p^{-1/D}$
%for large $p$s, and Eq. (\ref{eq:xi}) determines that $\alpha=1/D-1$

When $2\nu<2$ (when $1/3<D<1$) we are in the L\'{e}vy regime, and
$\alpha=2\nu=1/D-1$, so that
the leading order fractional L\'evy Edgeworth expansion of
$\tilde{w}_S(k)$ takes the
form:
\begin{equation}
{\tilde{w}_S}(k) \simeq e^{-a_\alpha|k|^\alpha}
\left\{\begin{array}{ll}\rule[5pt]{0mm}{10pt}
\mbox{ }\left[1+
\frac{\Gamma(1-\alpha/2)}{\Gamma(2-\alpha/2)2^2n^{2/\alpha-1}}|k|^2 
\right], &  \alpha>1 \\
\rule[5pt]{0mm}{15pt}
\left[1-
\frac{\Gamma^2(1-\alpha/2)}{\Gamma^2(1+\alpha/2)2^{2\alpha+1}n}|k|^{2\alpha}
\right], &  \alpha<1 \\
\rule[5pt]{0mm}{15pt}
1\mbox{ }, &\alpha=1
\end{array}\right.
\label{eq:fsK_Levy}
\end{equation}
where $a_\alpha \equiv {\Gamma(1-\alpha/2)}/{\Gamma(1+\alpha/2)2^\alpha}$ and
the leading term here agrees with
Eqs. (\ref{eq:fSk_correctionLevi})-(\ref{eq:CnCorrectionLevi}). In
the last case, corresponding to $D=1/2$, there is no correction term, since in
this case the single atom momentum distribution (in Eq.(\ref{eq:WeqP})) already
gives the Cauchy distribution, i.e., the $L_1(x)$ which is stable.

\subsection{Gaussian Regime}
\label{subSec:atomsGauss}
In order to find $W_{eq}(\mathcal{P})$, the PDF of the random variable
$\mathcal{P}\equiv \sum_{j=1}^{n}p_j/n^{1/2}$, we calculate numerically the
inverse Fourier transform of $\tilde{w}(k/n^{1/2})^n$ using Eq.
(\ref{eq:fkAtoms1}). In what follows we refer to this
result as \textit{the exact solution}, $W_{eq}(\mathcal{P})$.

\begin{figure}[tb]
\center{\includegraphics[width=0.7\textwidth]{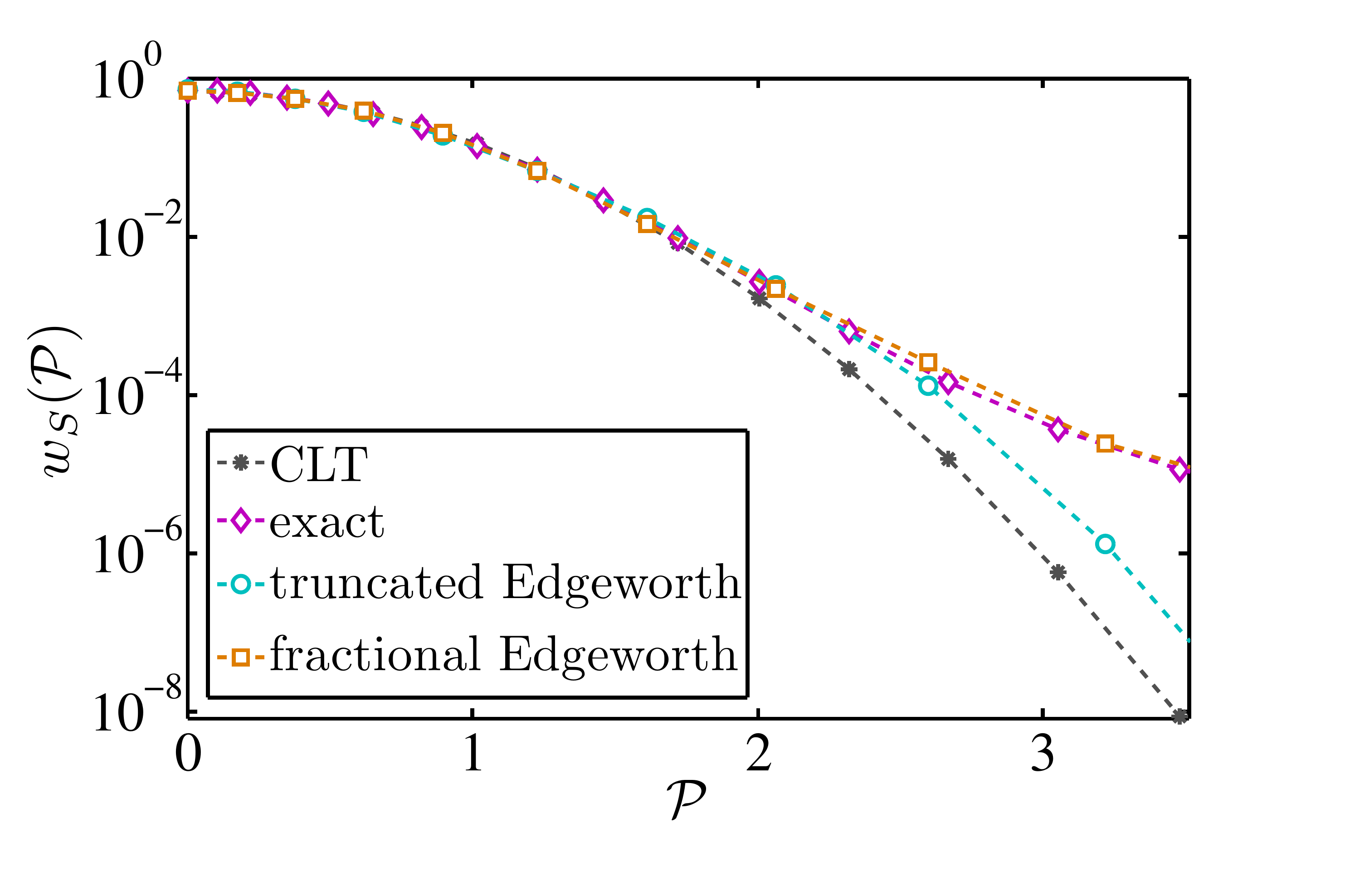}}
\caption{$w_S(\mathcal{P})$ for $D=1/6$ and $n=20$ drawn in a
semi-log scale. A comparison between the CLT, the exact solution, the truncated
Edgeworth expansion and the (leading order) fractional Gauss Edgeworth
expansion.}
\label{fig:ColdCompGauss1}
\end{figure}

\begin{figure}[tb]
\center{\includegraphics[width=0.7\textwidth]{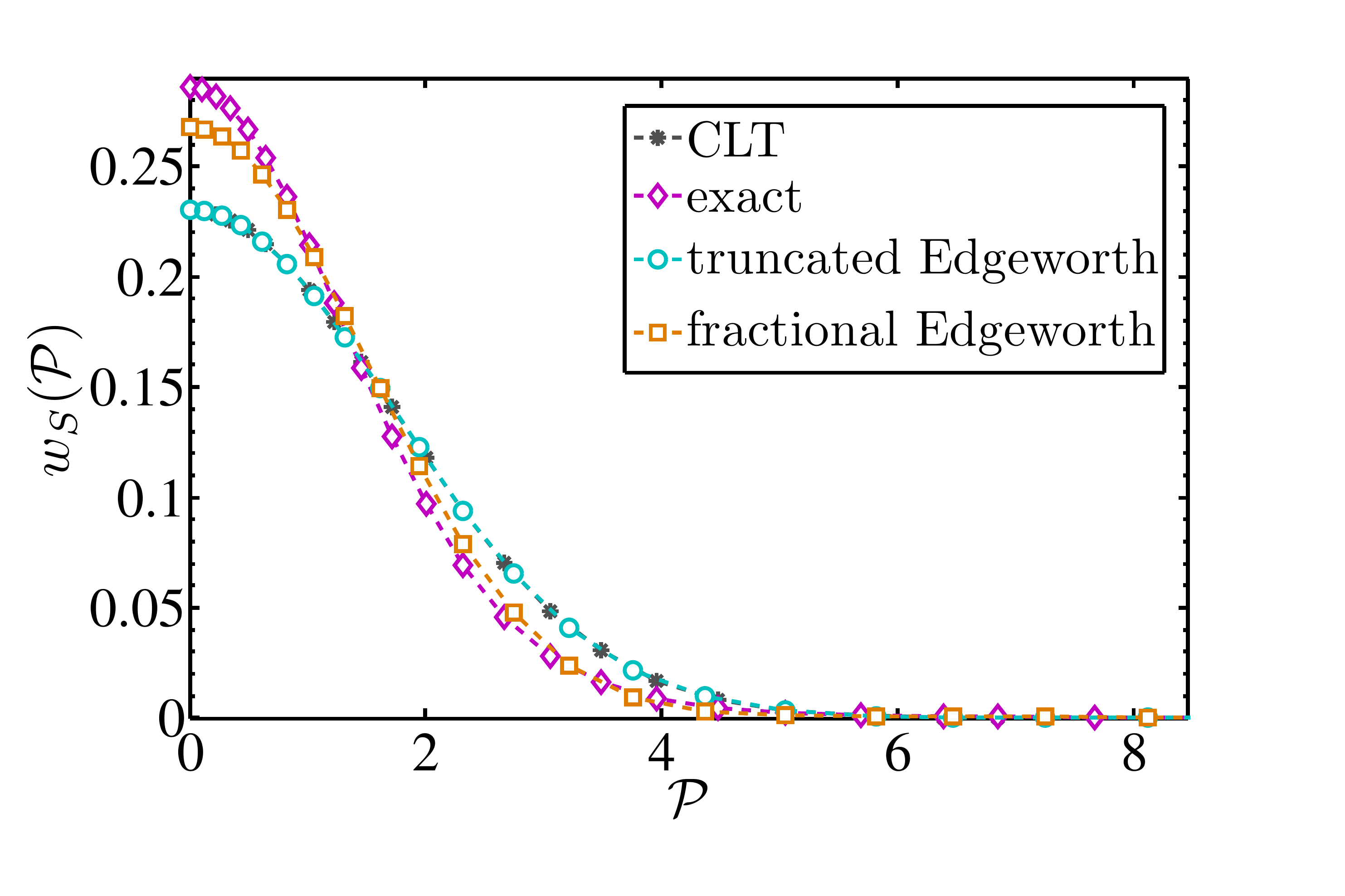}}
\caption{$w_S(\mathcal{P})$ for $D=0.3$ and $n=200$. A
comparison between the CLT, the exact solution, the truncated
Edgeworth expansion (coincides with the CLT for $D=0.3$) and
the (leading order) fractional Gauss Edgeworth expansion.}
\label{fig:ColdCompGaussCloseTo03}
\end{figure}

In the Gaussian regime, (even) moments exist only up to the highest integer that
is smaller than $\frac{1}{D}-1$. For example, for $1/5<D<1/3$ where even the
$4$th moment doesn't exist, the truncated Edgeworth reduces to the CLT. In
Fig. \ref{fig:ColdCompGauss1} and Fig. \ref{fig:ColdCompGaussCloseTo03} we
compare the CLT $Z_\sigma(\mathcal{P})=\frac{1}{\sqrt{2\pi\sigma^2}}
\exp(-\sigma^2\mathcal { P }^2/2)$, the exact solution $W_{eq}(\mathcal{P})$,
the truncated Edgeworth series $W^{(te)}(\mathcal{P})$, and the 
fractional Gauss Edgeworth expansion, $W^{(fge)}(\mathcal{P})$, for $D=1/6$
(corresponding to $\sigma^2=1/3$) and $D=0.3$ (corresponding to $\sigma^2=3$).
Using Eq. (\ref{eq:wSk_completeSeries}) without the non-analytic term (and
transforming back to $P$ space), the truncated Edgeworth expansion in this case
takes the form:
\begin{align}
W^{(te)}(\mathcal{P}) & = 
Z_{1/\sqrt{3}}(\mathcal{P}) + \frac{2(\sqrt{6})^5}{3\cdot 4!
n}T_{2,4}(\sqrt{6}\mathcal{P}) \nonumber\\
& =\sqrt{\frac{3}{2\pi}} e^{-3\mathcal{P}^2/2}  \bigg[1 +
\frac{3}{4n}\Big( 1-6\mathcal{P}^2 + 3\mathcal{P}^4  \Big) \bigg].
\label{eq:truncEdge}
\end{align}
Adding the first non-analytic term in Eq.
(\ref{eq:wSk_completeSeries}) (and transforming back to $\mathcal {P}$ space)
gives:
\begin{align}
W^{(fge)}(\mathcal{P})  &=  W^{(te)}(\mathcal{P})-\frac{2^3 3^3}{45n^{3/2}}
T_{2,5}(\sqrt{6}\mathcal{P}) \nonumber\\
& = W^{(te)}(\mathcal{P}) -
\frac{3}{5\pi n^{3/2}}\bigg[8+ 9\mathcal{P}^2(-3+\mathcal{P}^2) -
3\sqrt{6}\mathcal{P}\bigg(5-10\mathcal{P}^2 +
3\mathcal{P}^4\bigg)\textrm{Daw}\bigg(\sqrt{\frac{3}{2}}\mathcal{P}\bigg)
\bigg].
\label{eq:addedNaFirst}
\end{align}
By asymptotic expansion of the Dawson function for large $\mathcal{P}$ values,
we find that $W^{(fge)}(\mathcal{P})$ decays as $\sim 1/\mathcal{P}^6$ as
expected for $D=1/6$, since $W_{eq}(p)\sim p^{-1/D}$ in Eq. (\ref{eq:WeqP}).
As can be seen in Fig. \ref{fig:ColdCompGauss1}, the truncated
Edgeworth expansion fits the exact solution better than the CLT, but for the
tails of the density function this approximation breaks down.
Adding the non-analytic term to the Edgeworth series corrects this and the two
curves coincide even for the moderate $n=20$. As $D$ approaches $1/3$, the
fractional Gauss Edgeworth approximation converges to the exact solution for
higher $n$ values, while the truncated Edgeworth correction cannot recover the
exact solution behavior even for much higher $n$ values, and even the central
part of the truncated Edgeworth density function is significantly different from
the exact solution as can be seen, for example, in Fig.
\ref{fig:ColdCompGaussCloseTo03}
for $n=200$.

%\begin{figure}[tb]
%\center{\includegraphics[width=0.7\textwidth]{fig2.pdf}}
%\caption{A comparison between $n^*$ values for different D
%values, where $n^*$ measures the convergence of the truncated Edgeworth
%expansion (blue) and the \red{Hazut-expansion} (red) to the exact solution.}
%\label{fig:nStarGauss}
%\end{figure}

% \begin{figure}[tbh]
% \centering
% \subfigure[] { \label{fig:polyA}
% \includegraphics[scale=0.15]{fig1a_Kn1001elon  gated1.pdf}}
% %\newline%
% \centering
% \subfigure[]{\label{fig:polyB}
% \includegraphics[scale=0.15]{fig1b_Kn3001Localized1.pdf}}
% \centering
% \subfigure[]{\label{fig:polyC}
% \includegraphics[scale=0.15]{fig1c_Kn5001spread1.pdf}}
% \caption{Examples of plectonemic shapes for $3000bp$-long DNA
% with diameter $d=3.5nm$ and $\sigma = -0.07$: \textbf{(a)} unknotted chain
% \textbf{(b)} $3_1$ knot (a trefoil where we've colored the knotted portion of
% the chain in red) and \textbf{(c)} $5_{1}$ knot (the knot is spread over the
% entire chain)}
% \label{fig:somePlectonemes}
% \end{figure}

Since for $n\rightarrow\infty$ all the above PDFs coincide, and the higher 
$n$ is, the closer  the PDFs will be to each other, a good measure for
evaluating
the quality of these approximations is to calculate $n^*$ for which the
approximated PDF is close enough to the exact solution. We calculate $n^*$ as
the $n$ for which:
\begin{equation}
\int_{-\infty}^\infty\big( W_{ap}(\mathcal{P}) - W_{eq}(\mathcal{P}    \big)^2
d\mathcal{P} \leq \varepsilon_{cut},
\label{eq:measure}
\end{equation}
where $W_{ap}(\mathcal{P})$ corresponds to $W^{(te)}(\mathcal{P})$ or
$W^{(fge)}(\mathcal{P})$, and $\varepsilon_{cut}$ is a tunable threshold.

\begin{figure}[tb]
\center{\includegraphics[width=0.7\textwidth]{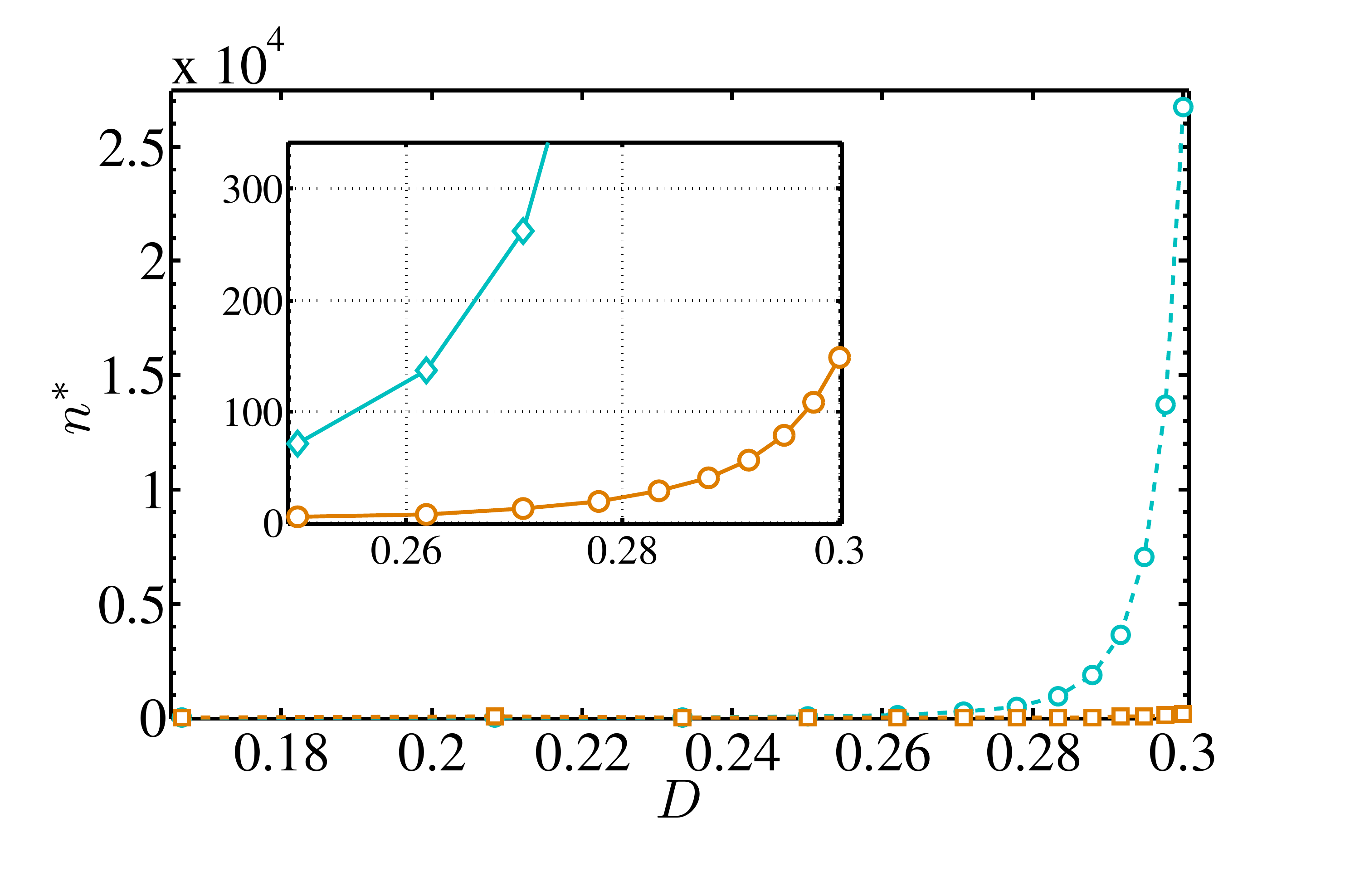}}
\caption{$n^*$ for different $D$ values in the Gaussian
regime, where $n^*$ measures the convergence of the truncated Edgeworth
expansion (blue) and the (leading order) fractional Gauss Edgeworth expansion
(orange) to the exact solution.  $\varepsilon_{cut}=3\cdot 10^{-4}$. 
The figure illustrates that the truncated Edgeworth does not work so well
compared to the fractional Gauss Edgeworth expansion.
In the inset we show that the leading order fractional Gauss Edgeworth indeed
has a slight increment in $n^*$ when $D$ approaches $1/3$.}
\label{fig:nStarGauss}
\end{figure}

In Fig. \ref{fig:nStarGauss} we examine the convergence of the approximate
PDFs to the exact solution for different $D$ values. As can be seen, whereas the
convergence of the truncated Edgeworth expansion becomes very slow as
$D\rightarrow 1/3$ (high values of $n^*$), the (first-term) fractional Gauss
Edgeworth approximation yields much faster convergence (smaller values of
$n^*$). The reason for this slow convergence of the truncated Edgeworth series
is that the Edgeworth series is an expansion around a Gaussian. The inverse
Fourier transform of the Edgeworth terms has the form
of $H_n(\mathcal{P}/\sigma)Z_\sigma(\mathcal{P})$ where $H_n$ is the Hermite
polynomial, and for large values of $\mathcal{P}$ the tails behavior is
controlled by the exponential decay which does not mimic the power-law decay of
the exact solution. The non-analytic expansion indeed decays according to the
exact
solution's power law, as we will now show. The inverse Fourier transform of the
non-analytic term is given by the integral:
\begin{equation}
%\begin{align}
\int_{0}^\infty dk\,k^\gamma e^{-\sigma^2 k^2/(2)}
%&
\cos(k\mathcal{P})  =
\frac{1}{2\sigma^{(\gamma+1)/2}}\Gamma(\gamma+1)e^{-\mathcal{P}^2/4\sigma^2}
%\nonumber\\&
 \bigg[D_{-(\gamma+1)}(-i\mathcal{P}/\sigma) +
D_{-(\gamma+1)}(i\mathcal{P}/\sigma)\bigg]
\label{eq:parab}
\end{equation}
%\end{align}
where $\gamma=1/D-1$, $D_{a}(z)$ is the parabolic-cylindric function
\cite{abramowitz1972handbook}, which for large $z$ goes to $\exp(-z^2/4)z^{a}$.
Substituting in this large $\mathcal{P}$ asymptotic behavior, the Gaussian term
cancels
and we are left with a power-law decay where $T_{2,\gamma}(\mathcal{P})\sim
1/\sigma^{\gamma+1}\Gamma(\gamma+1)\mathcal{P}^{-(\gamma+1)}$. As can be
seen in Fig. \ref{fig:ColdCompGaussCloseTo03} this addition of the
non-analytic term gives a pretty good approximation to the exact solution
suggesting that the power-law decay of the exact solution decays as the
expected $\mathcal{P}^{-(\gamma+1)}$.

The Edgeworth and the non-analytic corrections are still expansions around the
Gaussian, but as $D$ approaches $1/3$ we move from the Gaussian
regime towards the L\'{e}vy regime. As $D\rightarrow 1/3$ the convergence to the
exact solution becomes very slow, and only for extremely high $n$ do the PDFs
approach the exact solution. For all values of $D$, the (first-term)
non-analytic approximation yields faster convergence (smaller
value of $n^*$) than the truncated Edgeworth one due to the transition from
exponential to power-law decay.

For $D\rightarrow 1/3$ from below, even though the variance of $W_{eq}(p)$ is
finite, the convergence of the exact solution to a Gaussian is seen only for
extremely high $n$ values (see Fig. \ref{fig:nStarGauss}),
because the variance grows as $\sigma^2=D/(1-3D)$
which diverges at $D=1/3$.

\subsection{L\'{e}vy Regime}
\label{subSec:atomsLevy}
\begin{figure}[tb]
\center{\includegraphics[width=0.7\textwidth]{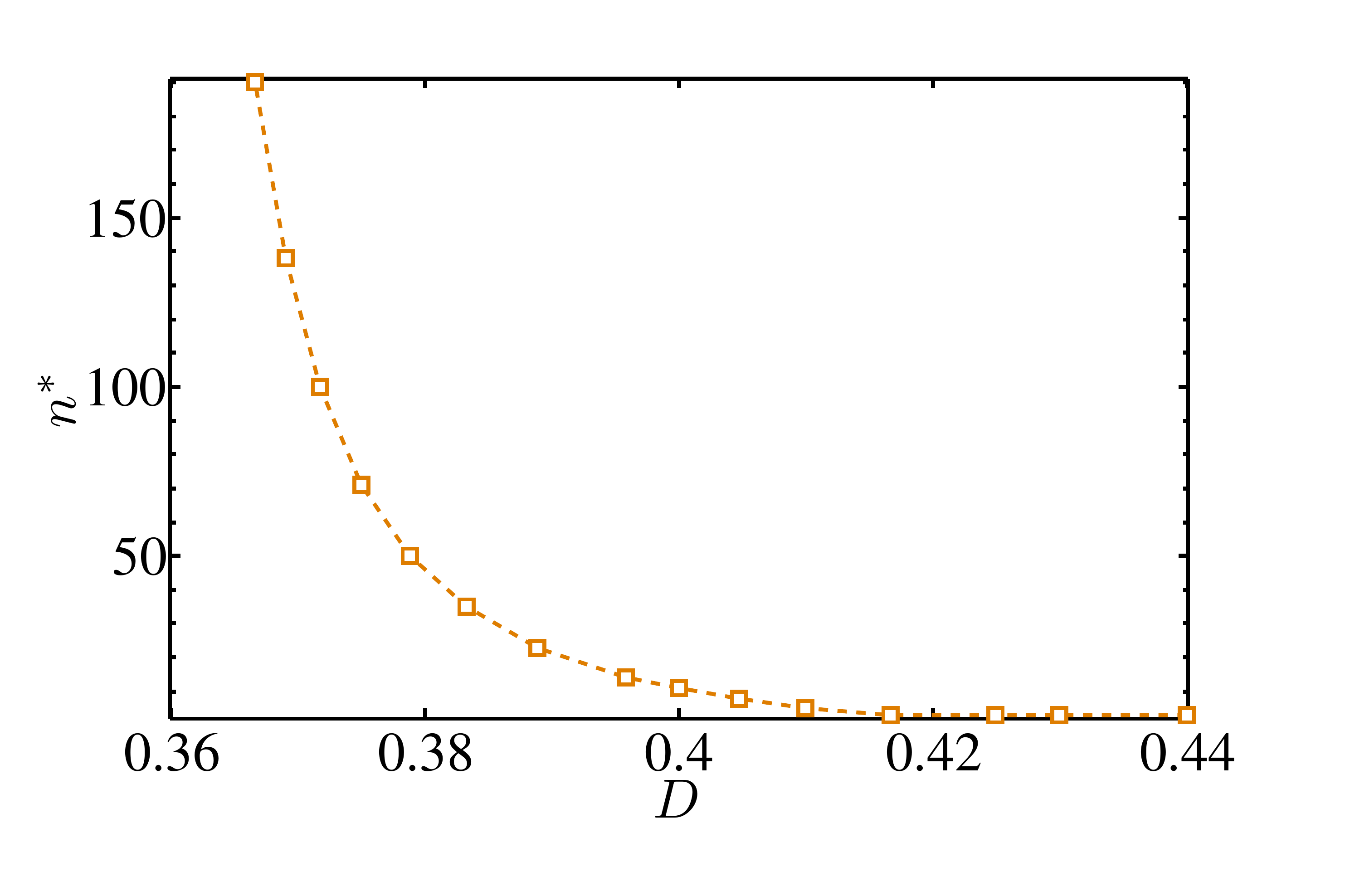}}
\caption{$n^*$ values for different $D$ values in the L\'{e}vy
regime, where $n^*$ measures the convergence of the leading order fractional
Edgeworth expansion
to the exact solution. $\varepsilon_{cut}=3\cdot 10^{-4}$.}
\label{fig:nStatLevi}
\end{figure}

For $1/3<D<1$, where the variance diverges, the basin of attraction of the PDF
is the L\'evy $\alpha$-stable density function, where for cold atoms, $\alpha
= \frac{1}{D}-1$. We will derive, therefore, the PDF of the random
variable $\mathcal{P}\equiv \sum_{j=1}^{n}p_j/n^{1/\alpha}$. As we have
discussed,
unlike the expansion in the Gaussian regime, the series in this regime is
entirely non-analytic in nature.
Although for large $n$ the exact solution tends to the L\'{e}vy
density function, when $D$ approaches $1/3$ (from above in the L\'{e}vy regime),
the $n$ needed for this convergence grows asymptotically. This is clearly shown
in Fig. \ref{fig:nStatLevi} where we've plotted $n^*$ (defined as above),
comparing the leading order fractional L\'evy Edgeworth and the exact solution.
Even though for large $n$ the PDF goes to the pure L\'{e}vy $\alpha$-stable
density function, when $D\rightarrow 1/3$ this convergence becomes very slow.

We will now show this slow convergence effect through the following example
cases. When $D=3/7$, corresponding to $\alpha=4/3$, $\gamma=2$, $a \simeq
1.178$ and $C_n=3/(4n^{1/2})$ and using Eqs.
(\ref{eq:fSk_correctionLevi})-(\ref{eq:wSx_correctionLevi}):
\begin{equation}
w_S(\mathcal{P}) \simeq
\frac{1}{a^{3/4}}\left[L_{\frac{4}{3}}(a^{-3/4}\mathcal{P})+
\frac{C_n}{a^{3/2}}T_{\frac{4}{3},2}(a^{-3/4}\mathcal{P}) \right].
\end{equation}
For $D=11/30$ which is much closer to $1/3$, we get the corresponding
$\alpha=19/11$, $\gamma=2$, $a \simeq 2.186$ and $C_n=11/(6n^{3/19})$
yielding:
\begin{equation}
w_S(\mathcal{P}) \simeq
\frac{1}{a^{11/19}}\left[L_{\frac{19}{11}}(a^{-11/19}\mathcal{P})+
\frac{C_n}{a^{22/19}}T_{\frac{19}{11},2}(a^{-11/19}\mathcal{P}) \right].
\label{eq:51}
\end{equation}

% \begin{figure}[tb]
% \center{\includegraphics[width=0.7\textwidth]{realFig9.pdf}}
% \caption{$w_S(\mathcal{P})$ for $D=3/7$ and $n=10$. A comparison
% between $L_\alpha(x)$, the exact solution and the
% leading order fractional L\'evy Edgeworth expansion.}
% \label{fig:ColdCompLevy}
% \end{figure}

\begin{figure}[tbh]
\centering
\subfigure[]{\label{fig:figS2histA}
\includegraphics[scale=0.27]{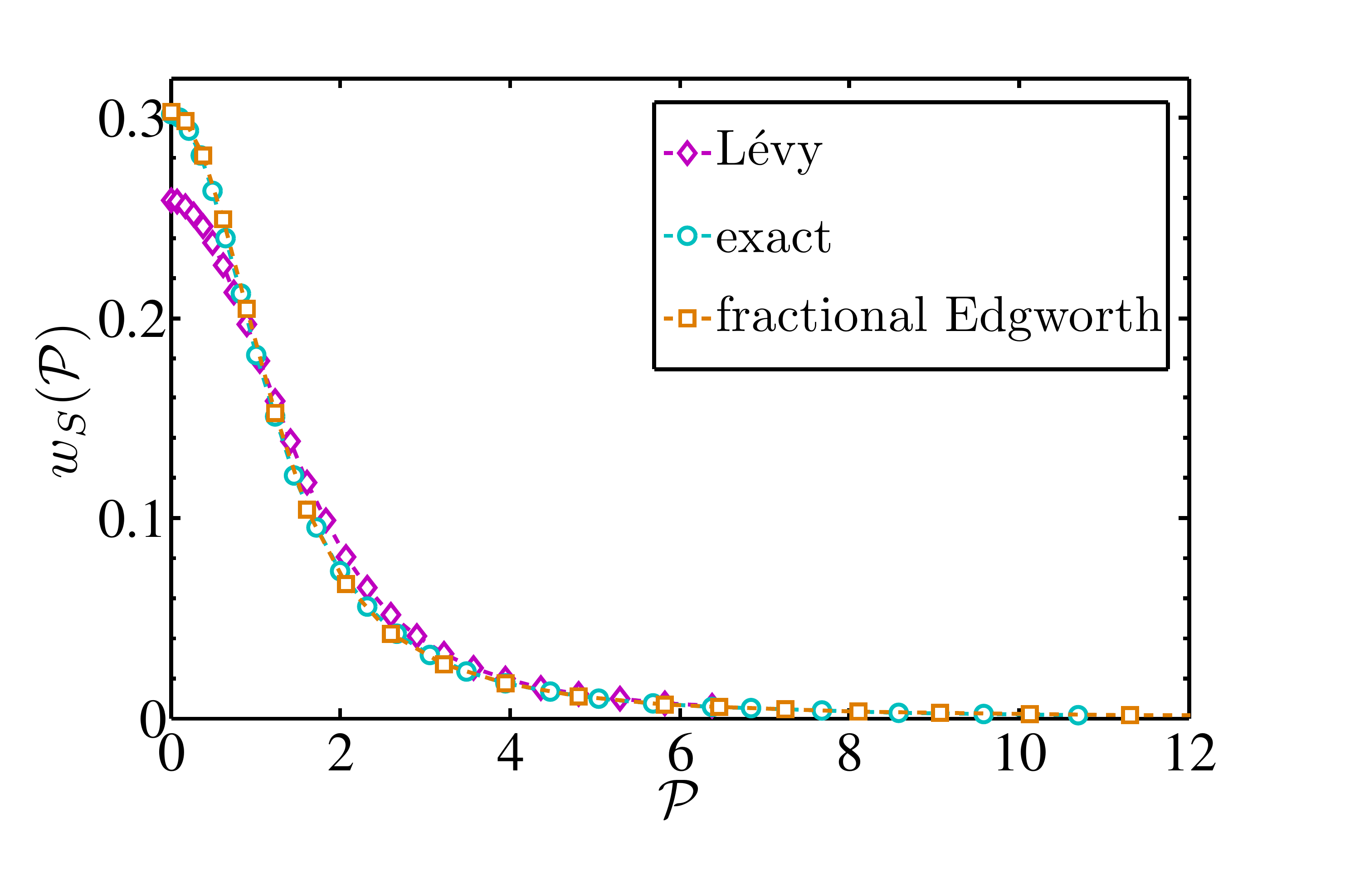}}
%\newline%
\centering
\subfigure[]{\label{fig:figS2histB}
\includegraphics[scale=0.27]{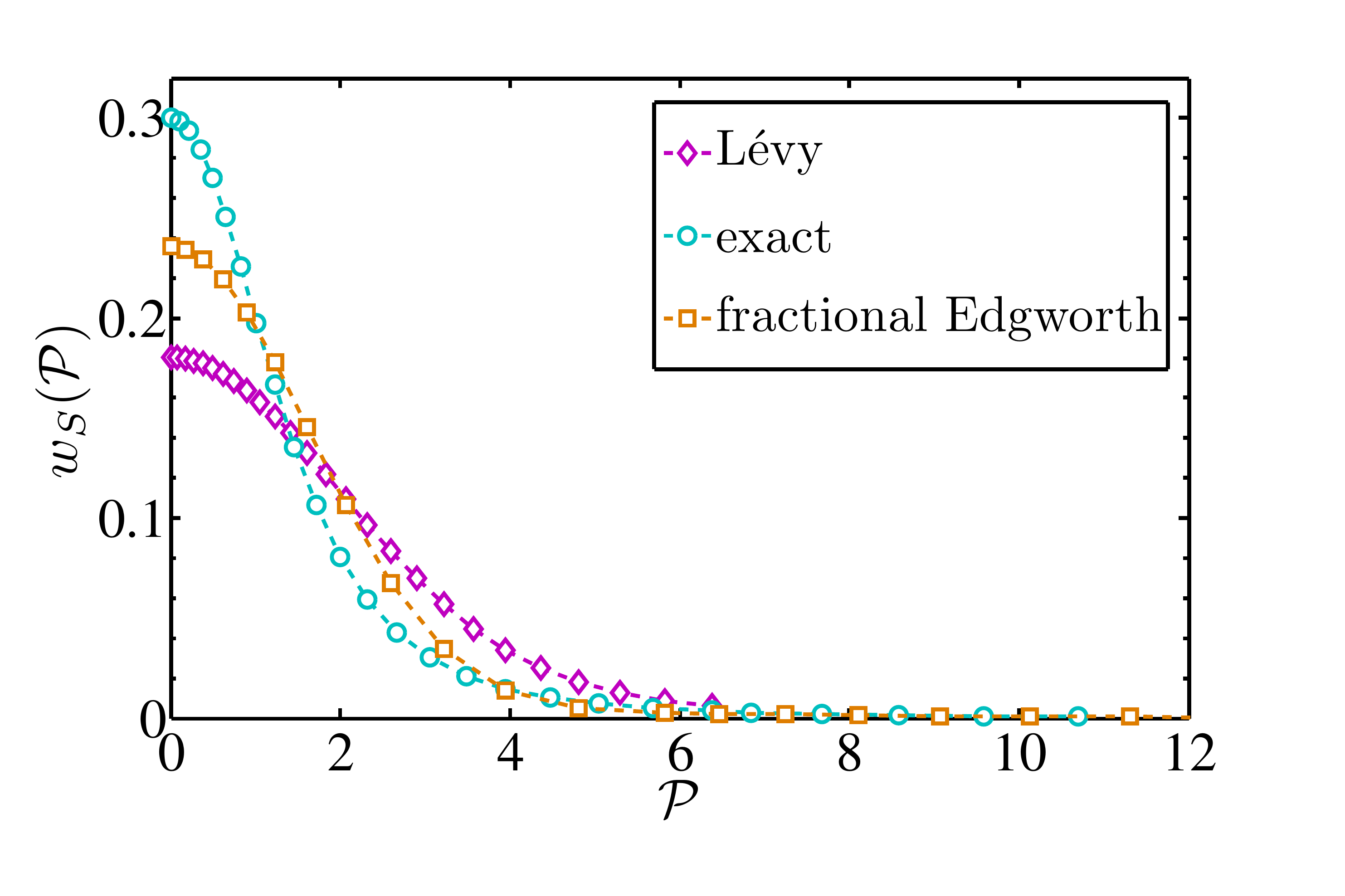}}
\newline%
\centering
\subfigure[]{\label{fig:figS2histB}
\includegraphics[scale=0.27]{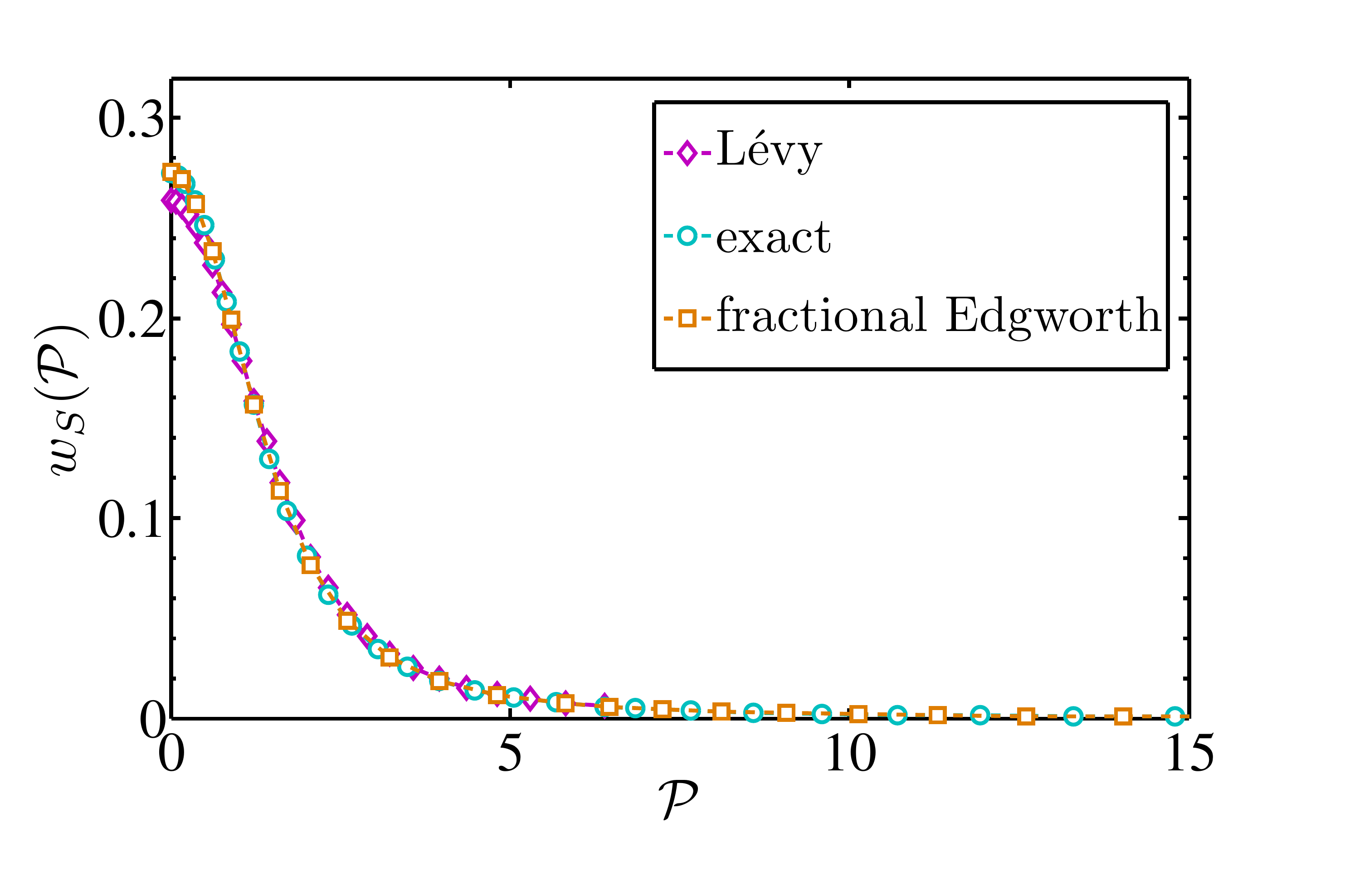}}
%\newline%
\centering
\subfigure[]{\label{fig:figS2histA}
\includegraphics[scale=0.27]{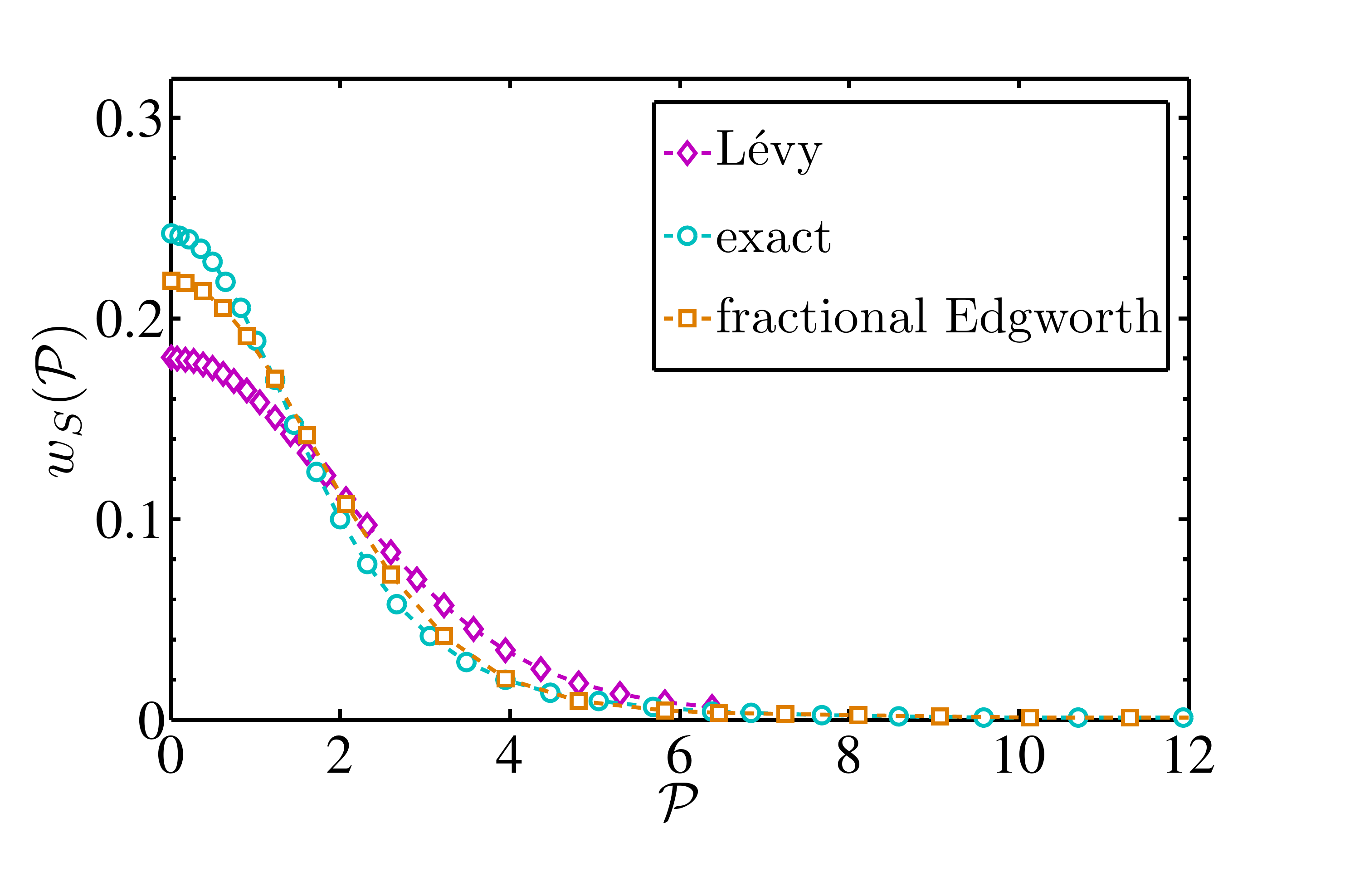}}
\newline%
\centering
\subfigure[]{\label{fig:figS2histA}
\includegraphics[scale=0.27]{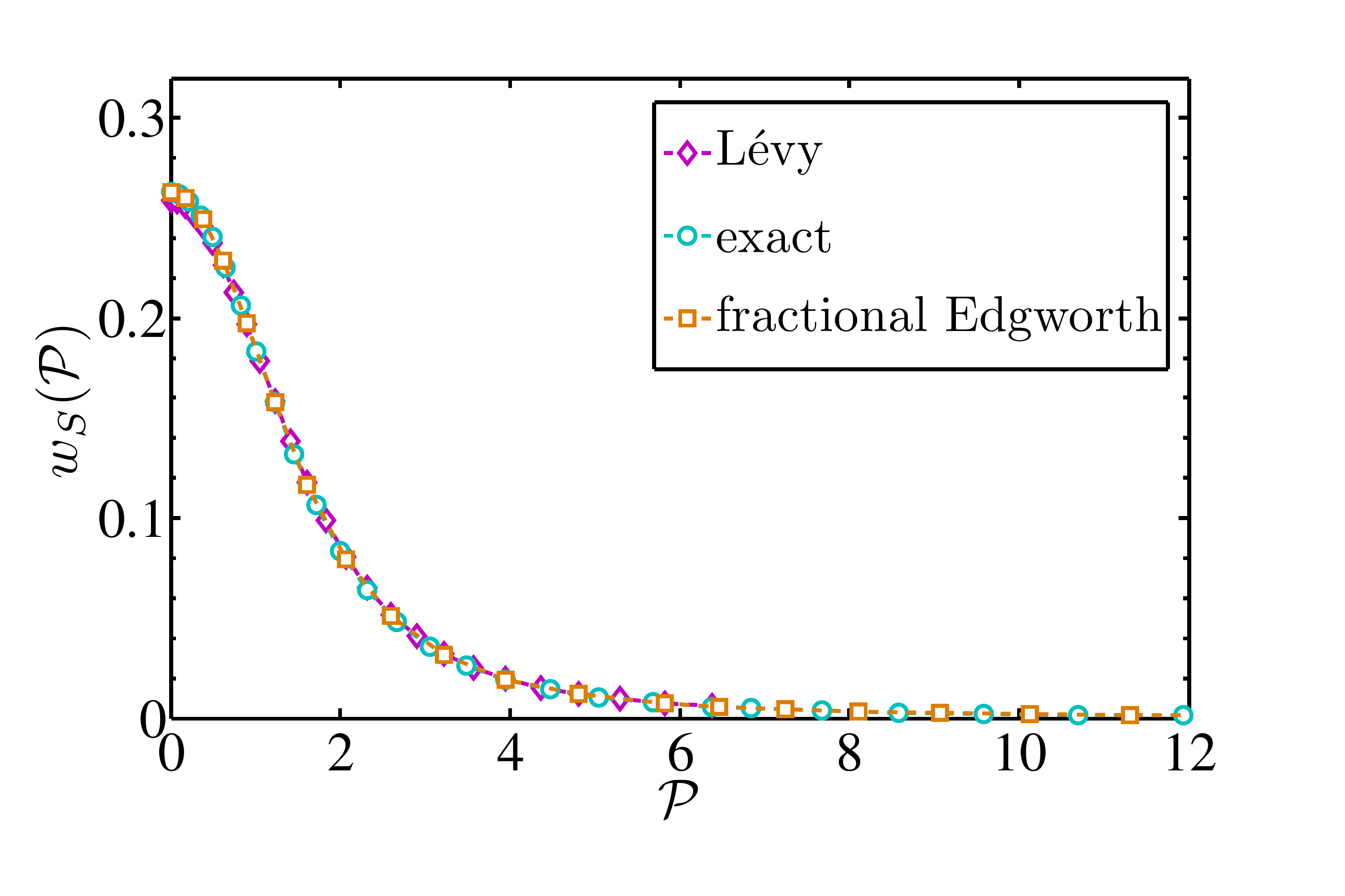}}
%\newline%
\centering
\subfigure[]{\label{fig:figS2histB}
\includegraphics[scale=0.27]{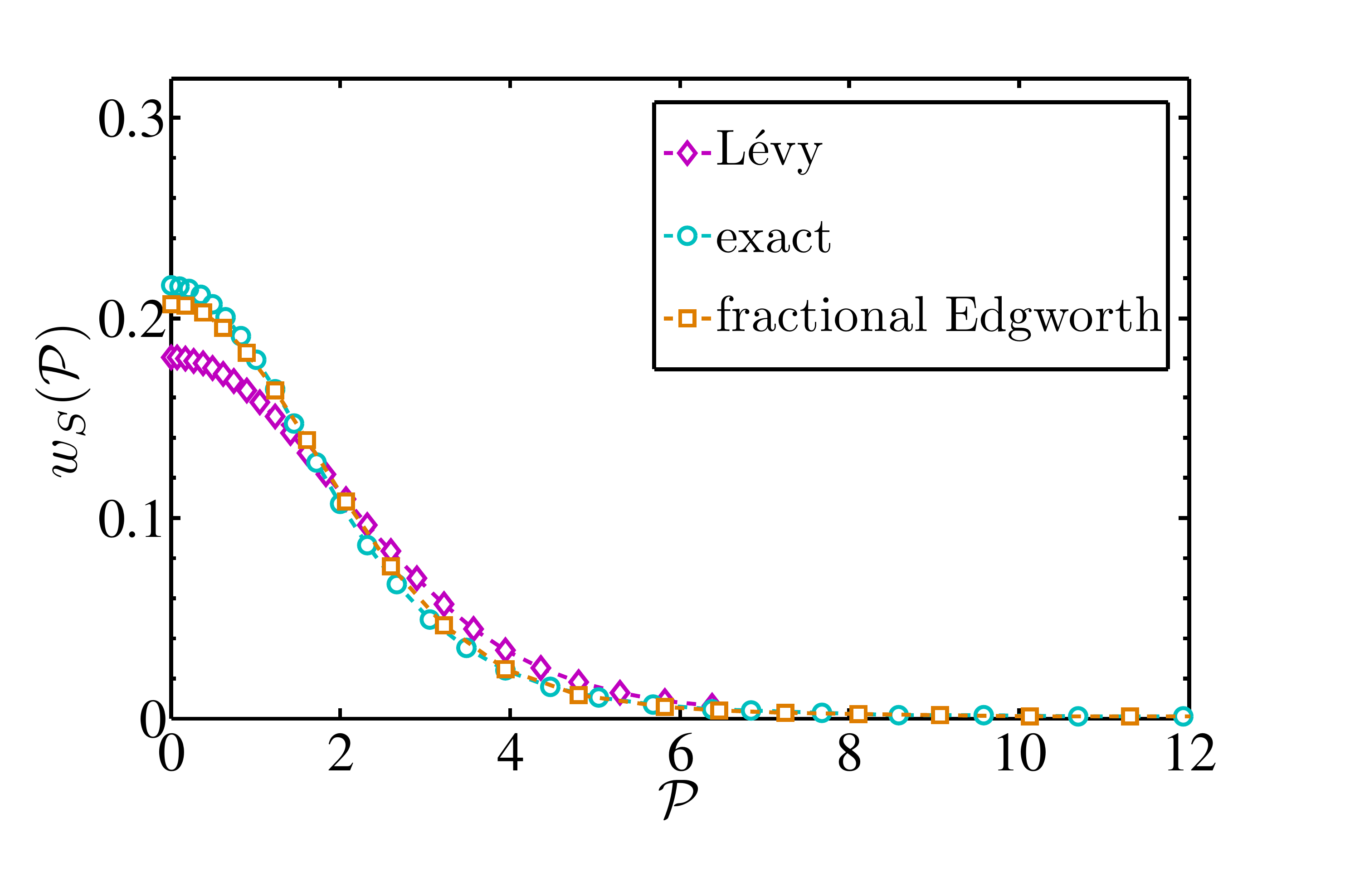}}
\caption{$w_S(\mathcal{P})$ for $D=3/7$, corresponding to $\alpha=4/3$ (left
hand side) and $D=11/30$, corresponding to $\alpha=19/11$ (right hand side) for
$n=10$ (first row), $n=100$ (second row) and
$n=1000$ (third row). A comparison between $L_\alpha(x)$, the exact solution and
the leading order fractional L\'evy Edgeworth expansion.}
\label{fig:fig9}
\end{figure}

In Fig. \ref{fig:fig9} we plot $w_S(\mathcal{P})$ for the above
examples for different $n$ values ($n=10,100,1000$) in order to compare the
convergence to the exact solution as $D$ approaches $1/3$.
It can be observed that for $D=3/7$, even for the moderate $n=10$ the correction
gives much better convergence to the exact solution, compared to
the $\alpha$-L\'{e}vy stable PDF. Increasing $n$, both the exact solution and
the leading term fractional Edgeworth approximation converge to the
$\alpha$-L\'{e}vy stable PDF. Nevertheless, the fractional L\'evy Edgeworth
approximation still approximates the exact solution better than the
$\alpha$-L\'{e}vy stable PDF. 
For $D=11/30$, however, the convergence is much slower. In the range of $n$
presented here, both the L\'evy density function and the corrected solution do
not coincide with the
exact solution, even though the fractional Edgeworth expansion gives a better
approximation
to the exact solution than the L\'evy density function. For much higher
$n$ values, however, the approximation Eq.(\ref{eq:51}) indeed coincides (up to
our numeric
accuracy) with the exact solution, as shown in Fig. \ref{fig:nStatLevi}.

\begin{figure}[tb]
\center{\includegraphics[width=0.7\textwidth]{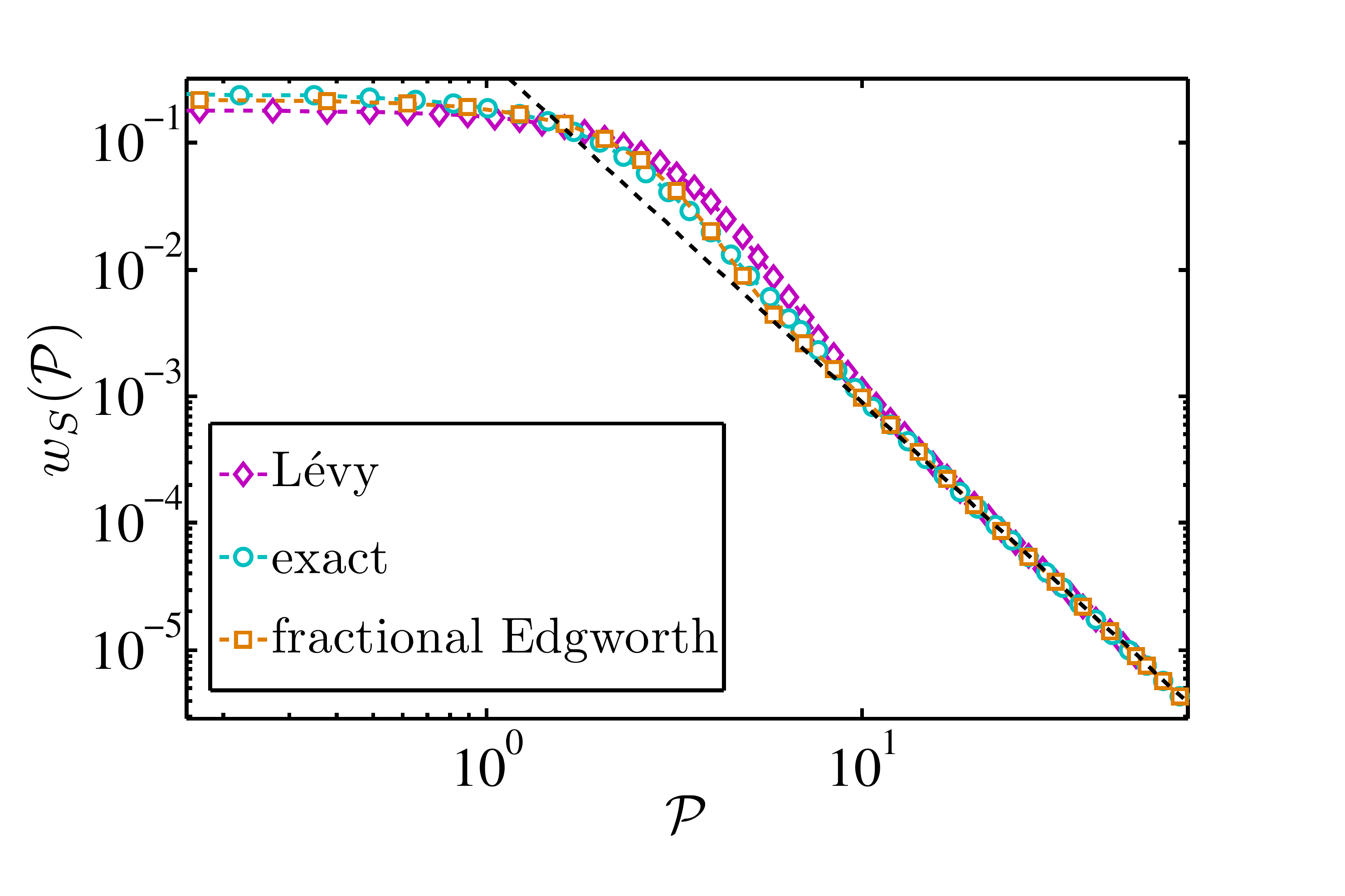}}
\caption{$w_S(\mathcal{P})$ for $D=11/30$ and $n=100$. A comparison
between $L_\alpha(x)$, the exact solution and the leading order fractional
L\'evy Edgeworth expansion in a log-log plot, in order to highlight the
power-low decaying behavior of the tails. The black dashed line has the expected
slope of $\alpha+1\simeq3$ for $D=11/30$, and all of these curves converge
to this slope for large $\mathcal{P}$.} 
\label{fig:semilog}
\end{figure}

In contrast to the Gaussian regime where the non-analytic term corrects the
tail's behavior from exponential decay to a power-law one, in the L\'{e}vy
regime, for high $\mathcal{P}$ values the density function already decays with
the same power-law as the L\'{e}vy, and the leading order fractional L\'evy
Edgeworth correction term takes care mostly of the center of the density
function. This
behavior of the tails is clearly shown in Fig. \ref{fig:semilog}, where we
plotted $w_S(\mathcal{P})$ for $D=11/30$ and $n=100$ in a semi-log plot,
\iffalse
where the tails of the L\'evy PDF, the fractional L\'evy Edgeworth approximation
and the exact solution have the same decay. 
\fi

\begin{figure}[tb]
\center{\includegraphics[width=0.7\textwidth]{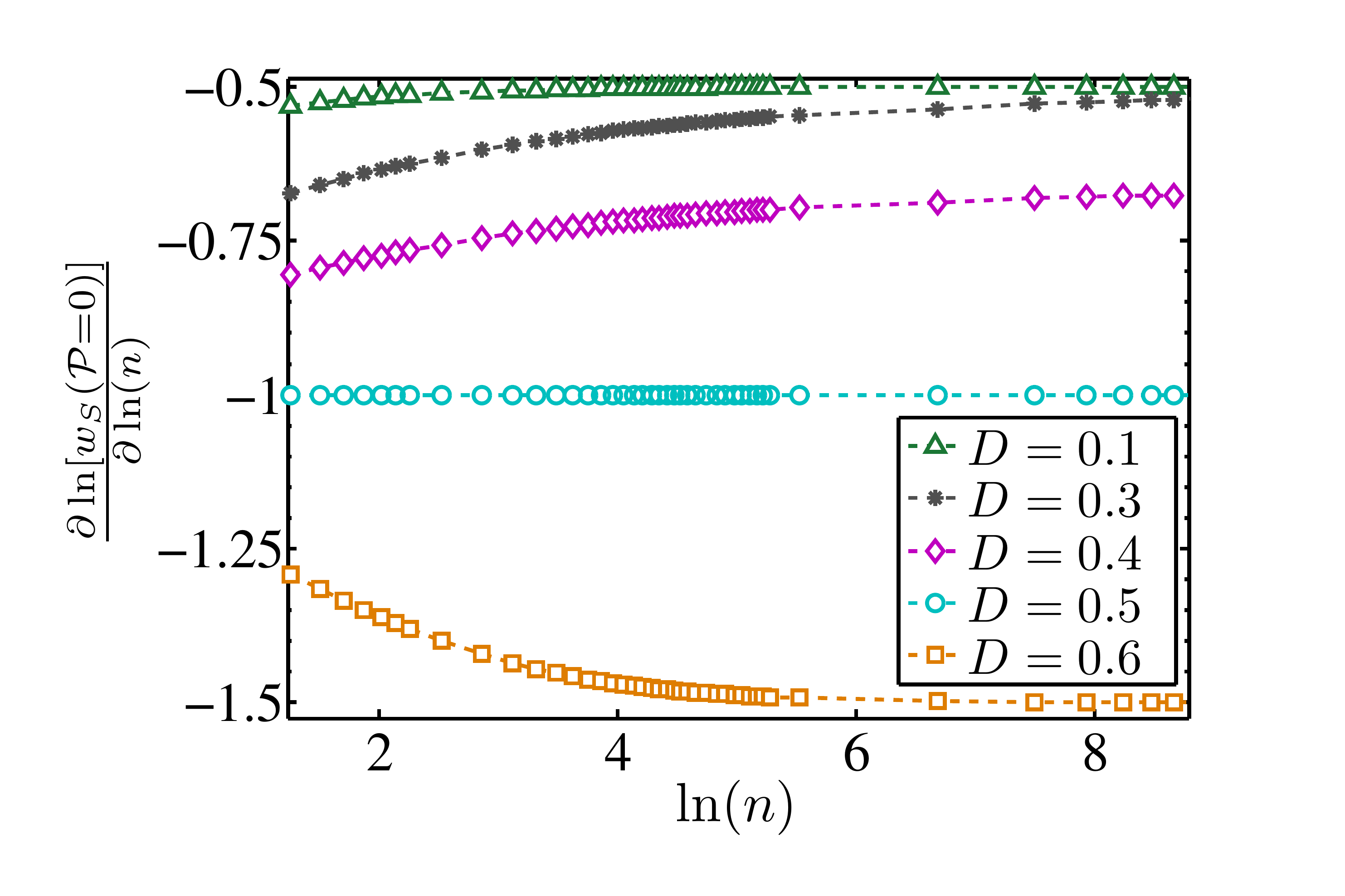}}
\caption{$\frac{\partial\log\left[ w_S(\mathcal{P}=0)
\right]}{\partial\log\left(n \right)}$ as a function of
$\log\left(n \right)$ for different $D$ values. each curve converges
asymptotically to its suitable $\alpha$. The curves corresponding to $D=0.1$
and $D=0.3$ tend asymptotically to $-0.5$ as expected in the Gaussian regime,
while the curve corresponding to $D=0.3$ converges for much higher $n$ values.
For $D$ equals $0.4$ ,$0.5$ and $0.6$ the curves tend asymptotically to
$-1/\alpha$, i.e., to $-2/3$, $-1$ and $-3/2$ respectively. In the L\'evy
regime, $D=0.5$ gives a pure L\'evy density function already for $n=1$. For
higher and lower $D$ values in this regime the convergence becomes slower, and
higher values of $n$ are needed for the asymptotic convergence.}
\label{fig:mantegna}
\end{figure}

A way to show the convergence of the central part of these PDFs to their basin
of attraction is to calculate the dependence of $w_S(\mathcal{P})$ on $n$ at
$\mathcal{P}=0$ \cite{mantegna1994stochastic}. It will be more instructive for
this purpose (in order to show the different
attractors of the Gaussian and L\'{e}vy regimes) to use the PDF of the sum
$\mathcal{P}=\sum_{i=1}^np_i$ instead of normalizing it by $n$ to the
appropriate power. For pure $Z_\sigma(\mathcal{P})$ and $L_\alpha(\mathcal{P})$
functions,
$w_S(\mathcal{P}=0)\propto n^{-1/\alpha}$ (where for $Z_\sigma(\mathcal{P})$ we
use $\alpha=2$). By plotting the dependence of $w_S(\mathcal{P}=0)$ on $n$ one
can
see how fast the density function converges to the stable density function.
In Fig. \ref{fig:mantegna} we plot $\partial\log\left( w_S(\mathcal{P}=0)
\right)/\partial\log\left(n \right)$ as a function of
$\log\left(n \right)$ which for large $n$ goes to $-1/\alpha$.
As can be seen, for $0<D<1/3$ the curves converge asymptotically to $-1/2$
while for $1/3<D<1$ each curve converges to its appropriate $-1/\alpha =
D/(D-1)$. Also, one may observe that as $D$ approaches $1/3$ from both sides,
the convergence of the curves become much slower.

\section{Summary}
\label{sec::summary} 
In this article we have generalized the classical Edgeworth expansion for finite
$n$ to PDFs which converge to the $\alpha$-stable L\'evy density functions. In order
to do this we used a generalized Fourier series including fractional powers of
$k$ and showed that the inverse Fourier transform of this series may be written
by means of a series of fractional derivatives of the L\'evy PDF and its conjugate,
$R_{\alpha}(x)$ (Eq. (\ref{eq:p_conjLevi})). This generalization is shown to
be universal since it also gives the classical Edgeworth series for PDFs in the
Gaussian regime when all the moments exist, and the fractional Gauss Edgeworth
expansion developed by Lam, et al \cite{lam2011corrections}. for PDFs with
finite variance but
diverging moments.

For the correction terms we introduced a new family of special functions,
$T_{\alpha,\gamma}(x)$ Eq. (\ref{eq:T_ag1}), for which  the Gaussian, the
L\'evy
and the Hermite-Gauss functions are special cases. We also represented
these functions as  $H$-Fox functions via the Mellin-Barnes integral.
\iffalse
, and using
these identities we related $T_{\alpha,\gamma}$s with different indices to each
other, enabling us to obtain the fractional derivatives of the L\'evy and the
conjugate L\'evy PDFs by shifting the indices of their $H$-Fox functions
presentations.
\fi 
We investigated the behavior of these functions in the context of our
correction
terms (for specific values  of $\alpha$ and $\gamma$).
\iffalse
but further study of
these
functions is required for a full understanding and usefulness of this family of
functions.
\fi

We have applied our results to the sum of momenta of cold atoms, and showed
that taking even only the first correction term of our fractional series
(leading term approximation) already gives much better matching to the
exact solution for small values of $n$. At the transition between Gauss to
L\'evy behaviors, we have found very slow convergence to the asymptotic result.

% \begin{figure}[tbh]
% \centering
% \subfigure[] { \label{fig:Fig1A_alpha05}
% \includegraphics[scale=0.26]{fig6_T_AlphaGamma05.pdf}}
% \newline%
% \centering
% \subfigure[]{\label{fig:Fig1B_alpha13}
% \includegraphics[scale=0.26]{fig7_T_AlphaGamma13.pdf}}
% \caption{(a) ??? (b) ??? (c) ???}
% \label{fig:Fig1}
% \end{figure}
% 
% 
% \begin{figure}[tbh]
% \label{fig:somePlectonemes}
% \centering
% \subfigure[] { \label{fig:Fig2A}
% \includegraphics[scale=0.26]{fig9_T_DifAlphaForGamma05.pdf}}
% \subfigure[]{\label{fig:Fig2B}
% \includegraphics[scale=0.26]{fig6b_T_AlphaGamma05_changeFromPosToNeg.pdf}}
% \caption{(a) ??? (b) ??? (c) ???}
% \label{fig:Fig2}
% \end{figure}

\section{Acknowledgments}
\label{sec:acknowledgments}
This work is supported by the Israel Science Foundation (ISF).

%%%%%%%%%%%%%%%%%%%%%%%%%%%%%%%%%%%%%%%%%%%%%%%%%%%%%%%%%%%%%%%%%%%%%%%%%%%%%%%%
%%%%%%%%%%%%%%%%%%%%%%%%%%%%%%%%%%%%%%%%%%%%%%%%%%%%%%%%%%%%%%%%%%%%%%%%%%%%%%%%
%%%%%%%%%%%%%%%%%%%%%%%%%%%%%%%%%%%%%%%%%%%%%%%%%%%%%%%%%%%%%%%%%%%%%%%%%%%%%%%%
\appendix
\section{$T_{\alpha,\gamma}(x)$ as the $H$-Fox Function} 
\label{app::hFox}

\subsection{The $H$-Fox function} 
\label{appsub::hFox}

Fox \cite{fox1961g,braaksma1936asymptotic} defined the $H$-function by  
\begin{equation}
H^{m,n}_{p,q}(z)=\frac{1}{2\pi i} \int_{L}\chi(s)z^{s}ds, 
\label{eq:integralH}
\end{equation}
where $L$ is a path in the complex plane $\mathbb{C}$ to be described later,
$z=\exp(\log|z|+i \;\arg z)$ and the integral density $\chi(s)$ is given
by  \begin{equation} 
\chi(s)=\frac{A(s)B(s)}{C(s)D(s)}=\frac{\prod_{j=1}^m\Gamma(b_j-B_js)\prod_{j=1}
^n\Gamma(1-a_j+A_js)}{\prod_{j=m+1}^q\Gamma(1-b_j+B_js)\prod_{j=n+1}
^p\Gamma(a_j-A_js)},
\label{eq:sets}
\end{equation}
$n$,$m$,$p$,$q$ are integers satisfying 
\begin{equation}
\nonumber 0\leq n\leq p, \quad 1\leq m\leq q ,
\end{equation}
$A_j,B_j$ are positive numbers and $a_j,b_j $ are in general
complex numbers. When no elements appear in one of the multiplications in Eq.
\ref{eq:sets} one gets an empty product which is taken to equal unity: 
\begin{equation}
\nonumber
m=0\rightarrow A(s)=1,\quad n=0\rightarrow B(s)=1,\quad m=q
\rightarrow C(s)=1,\quad n=p\rightarrow D(s)=1.
\end{equation} 
Since $H^{m,n}_{p,q}(z)$ depends on the sets $\{a_i, A_i\}$ and $\{b_i, B_i\}$,
a common notation for $H^{m,n}_{p,q}(z)$ is:
\begin{equation}
 H^{m,n}_{p,q}(z)\equiv
 H^{m,n}_{p,q} \left[ {z} \bigg| \begin{array}{cc} {(a_j,A_j)_{j=1,\ldots,p}} \\
{(b_j,B_j)_{j=1,\ldots,q}} \end{array} \right] \\
\end{equation}

This representation of the $H$-Fox function via an integral involving products and
ratios of Gamma functions is known as the Mellin-Barnes integral
\cite{marichevhandbook,paris2001asymptotics}.
The singular points of the kernel $\chi(s)$ are the poles of the
Gamma functions in $A(s)$ and $B(s)$, which are assumed to not coincide.
Denoting the sets of poles by $P(A)$ and $Q(B)$ respectively, $P(A)\cap
Q(B)=\emptyset$. The conditions for the existence of the $H$-Fox function can be
determined by examination of the convergence of the integral
in Eq. (\ref{eq:integralH}), which depends on the selection of the contour $L$
and on certain relations between the parameters $\{a_i,A_i\}$ where $i=1,\ldots,
p$ and $\{b_j,B_j\}$ where $j=1, \ldots, q$.
The contour $L$ in Eq.(\ref{eq:integralH}) can be chosen as the contour in which
all poles of $P(A)$ lie to its
right and all poles of $Q(B)$ to lie its left while further $L$ runs from 
$c-i\infty$ to $c+i\infty$. Other kinds of Barnes-contours are also possible
(see e.g., \cite{braaksma1936asymptotic, mathai2009h}).
\iffalse
The contour $L$ in Eq.(\ref{eq:integralH}) can be chosen by each as follows:
of $P(A)$ lies to its right  and all the poles of $Q( leaves all the poles of $P(A)$ to its right, and all the poles of $Q(B)$ to its       \begin{enumerate}[label={(\roman*)}]
 \item  $L=L_{-i\infty,+i\infty}$ chosen in such a way to go from
$-i\infty$ to $+i\infty$ leaving to the right all the poles of
$P(A)$, namely the poles $s_{j,k}=(b_j+k)/B_j$ where $j=1,2,\ldots,m \quad
k=0,1,\ldots$  of
the $\Gamma$ functions entering $A(s)$, and to the left all the poles of $Q(B)$:
$s_{j,l}=(a_j-1-l)/B_j$ where $j=1,2,\ldots,n \quad l=0,1,\ldots$ of the
$\Gamma$ functions entering $B(s)$
\item$L=L_{+\infty}$ is a loop beginning and ending at
$+\infty$ and encircling all the poles of $P(A)$ once in the negative direction,
but none of the poles of $P(B)$.
\item$L=L_{\-\infty}$ is a loop beginning and ending at
$-\infty$ and encircling once in the positive direction all the poles of $Q(B)$,
but none of the poles of $P(A)$.
\end{enumerate}
\fi

It was shown by Braaksma \cite{braaksma1936asymptotic} that the
Mellin-Barnes integral in Eq. \ref{eq:integralH} makes sense and defines an
analytic function of $z$ in
the following two cases:
\begin{enumerate}[label={(\roman*)}]
 \item  \begin{equation} 
\mu=\sum_{i=1}^qB_i-\sum_{j=1}^pA_j>0,\quad \forall z\not=0
\end{equation}.
\item\begin{equation}
 \mu=0 \quad and \quad 0\leq |z|\leq \delta\quad where\quad
\delta=\prod_{i=1}^q(B_i)^	{B_i} \prod_{j=1}^p(A_j)^{-A_j}.
\end{equation}

\end{enumerate}

The convergence and asymptotic expansions (for $z\rightarrow 0$
and $z\rightarrow\infty$) are determined by applying the residue theorem at the
poles (which are by assumption simple poles) of the Gamma functions in $A(s)$ and
$B(s)$.

The $H$-Fox function has a few properties which are important to our purposes:
\begin{enumerate}[label={(\roman*)}]
 \item The $H$-Fox function is symmetric in the pairs
$(a_1,A_1),\ldots,(a_n,A_n)$, likewise $(a_{n+1},A_{n+1}),\ldots,(a_p,A_p)$; in
$(b_1,B_1),\ldots,(b_m,B_m)$ and in $(b_{m+1},B_{m+1}),\ldots,(b_q,B_q)$.

\item If one of the $(a_j,A_j)$ $j=1,\ldots,n$, is equal to one of the
$(b_j,B_j)$, $j=m+1,\ldots,q$ or one of the pairs $(a_j,A_j)$, $j=n+1,\ldots,p$
is equal to one of the $(b_j,B_j)$, $j=1,\ldots,m$ then the $H$-Fox function
reduces to a lower order $H$-function, namely, $p$ and $q$, and $n$ (or $m$)
decrease by unity. Provided that $n\geq1$ and $q>m$ we have:
\begin{equation}
 H^{m,n}_{p,q} \left[ z \bigg| \begin{array}{cc} {(a_j,A_j)_{1,p}} \\
{(b_j,B_j)_{1,q-1}(a_1,A_1)} \end{array} \right]=H^{m,n-1}_{p-1,q-1} \left[ z
\bigg| \begin{array}{cc} {(a_j,A_j)_{2,p}} \\ {(b_j,B_j)_{1,q-1}} \end{array}
\right],
\end{equation}
\begin{equation}
 H^{m,n}_{p,q} \left[ z \bigg| \begin{array}{cc} {(a_j,A_j)_{1,p-1}(b_1,B_1)} \\
{(b_1,B_1)(b_j,B_j)_{2,q}} \end{array} \right]=H^{m-1,n}_{p-1,q-1} \left[ z
\bigg| \begin{array}{cc} {(a_j,A_j)_{1,p-1}} \\ {(b_j,B_j)_{2,q}} \end{array}
\right]
\end{equation}
\item
\begin{equation}
 z^{\sigma}H^{m,n}_{p,q} \left[ {z} \bigg| \begin{array}{cc} {(a_j,A_j)_{1,p}}
\\ {(b_j,B_j)_{1,q}} \end{array} \right]=  H^{m,n}_{p,q}(z)=
 H^{m,n}_{p,q} \left[ {z} \bigg| \begin{array}{cc} {(a_j+\sigma A_j,A_j)_{1,p}}
\\ {(b_j+\sigma B_j,B_j)_{1,q}} \end{array} \right]
\end{equation}
\item
\begin{equation}
 \frac{1}{c}H^{m,n}_{p,q} \left[ {z} \bigg| \begin{array}{cc} {(a_j,A_j)_{1,p}}
\\ {(b_j,B_j)_{1,q}} \end{array} \right]=  H^{m,n}_{p,q}(z)=
 H^{m,n}_{p,q} \left[ {z} \bigg| \begin{array}{cc} {(a_j,cA_j)_{1,p}} \\
{(b_j,cB_j)_{1,q}} \end{array} \right],\quad c>0
\end{equation}
\item
Another useful and important formula for the $H$-Fox function is
\begin{equation}
 H^{1,1}_{2,2} \left[ z \bigg| \begin{array}{cc} {(a_j,A_j)_{1,p}} \\
{(b_j,B_j)_{1,q}} \end{array} \right]=H^{1,1}_{2,2} \left[ \frac{1}{z} \bigg|
\begin{array}{cc} {(1-b_j,B_j)_{1,q}} \\ {(1-a_j,A_j)_{1,p}} \end{array}
\right],
\label{eq:iden}
\end{equation}

\end{enumerate}

This last relation enables us to transform an $H$-Fox function with $\mu<0$
and argument $z$ to one with $\mu>0$ and argument $1/z$.

\subsection{Representation of $T_{\alpha,\gamma}(x)$} 
\label{appsub:tAlphaGamma}

In order to represent $T_{\alpha,\gamma}(x)$ by the $H$-Fox function first we
express
$T_{\alpha,\gamma}(x)$  as a Mellin-Barnes integral. $T_{\alpha,\gamma}(x)$
is
defined as the following inverse Fourier transform :
\begin{equation}
\frac{1}{\pi} \int_0^\infty \cos(kx)
\exp(-k^\alpha)k^\gamma dk =
\Re\left\{\frac{1}{\pi}\int_0^\infty \cos(kx)
\exp(-ikx-k^\alpha)k^\gamma dk\right\}. 
\end{equation}
Using the Mellin-Barnes representation of $\exp(-ikx)$:
\begin{equation}
\exp(-ikx)=\frac{1}{2\pi i} \int_L \Gamma(s)(ikx)^{-s}ds
\label{eq:MBexp}
\end{equation}
(where $L$ is a loop in the complex $s$ plane that encircles the poles of
$\Gamma(s)$ in the positive sense with end-points at infinity at $\Re(s)<0$),
we get:    
\begin{align}
T_{\alpha,\gamma}(x)=& \Re\bigg\{ \frac{1}{\pi}\int_0^{\infty}
\exp(-k^{\alpha})k^{\gamma}\bigg[ \frac{1}{2 \pi i} \int_L
\Gamma(s)(ikx)^{-s}ds\bigg]dk\bigg\} 
\nonumber\\
&= \frac{1}{\pi} \frac{1}{2 \pi i} \int_L \Gamma(s) \Re \bigg\{ \int_0^{\infty}
\exp(-k^{\alpha})k^{\gamma}(ik)^{-s}dk \bigg\} x^{-s}ds.
\end{align}
The term in the brackets can be written as:
\begin{align}
\Re\bigg\{ \int_0^{\infty} \exp(-k^{\alpha})k^{\gamma}(ik)^{-s}dk
\bigg\}&=\Re\bigg\{ i^{-s} \frac{1}{\alpha}\Gamma(\frac{1+ \gamma -s}{\alpha})
\bigg\}
\nonumber\\
&=\frac{1}{\alpha}\cos(\frac{\pi s}{2})\Gamma(\frac{1+ \gamma
-s}{\alpha})=\frac{1}{\alpha}\sin(\frac{\pi}{2}(s+1))\Gamma(\frac{1+ \gamma
-s}{\alpha})
\nonumber\\
&=\frac{\pi}{\alpha}\frac{\Gamma(\frac{1+ \gamma
-s}{\alpha})}{\Gamma(\frac{1-s}{2})\Gamma(\frac{1+s}{2})},
\end{align}
which finally gives:
\begin{equation}
T_{\alpha,\gamma}(x)= \frac{1}{\alpha}\frac{1}{2\pi i} \int_L
\frac{\Gamma(s)\Gamma(\frac{1+\gamma-s}{\alpha})}{\Gamma(\frac{1-s}{2}
)\Gamma(\frac{1+ s}{2})}x^{-s}ds.
\label{eq:tagS}
\end{equation}
 This Mellin-Barnes integral can be written in terms of a $H$-Fox function,
following Eqs. (\ref{eq:integralH}) and (\ref{eq:sets}):
\begin{equation}
T_{\alpha,\gamma}(x)=\frac{1}{\alpha}H^{1,1}_{2,2} \left[ x \bigg|
\begin{array}{cc} {(1-\frac{1+\gamma}{\alpha},\frac{1}{\alpha}),(\frac{1}{2},
\frac{1}{2})} \\ {(0,1),(\frac{1}{2},\frac{1}{2})} \end{array} \right].
\label{eq:TFox1}
\end{equation}

\subsection{Asymptotic Expansion of $T_{\alpha,\gamma}(x)$} 
\label{appsub:AsymptototicExpansion}
The simple poles of $\Gamma(s)$ and $\Gamma(\frac{1+ \gamma -s}{\alpha})$ in Eq.
(\ref{eq:tagS}) are given by the disjoint sets of points:
\begin{align}
\nonumber
&P(s)=\{s_{\nu}=1 + \gamma+ \alpha \nu, \quad \nu=0,1,\cdots\}
\nonumber\\
&Q(s)=\{s_{\nu}=-\nu, \quad \quad\quad\quad \nu=0,1,\cdots\}.
\nonumber 
\end{align}
We distinguish between the following two cases:

\begin{enumerate}[label={(\roman*)}]
\item $x\rightarrow \infty$: Choosing the contour ${L}$ in Eq. (\ref{eq:tagS}) as ${L}={L}_{-i\infty,+i\infty}$ and closing the contour
to the right by a semi-circle of radius $R \rightarrow \infty$, we obtain the
large $x$ series asymptotic expansion:
\begin{align}
&H^{1,1}_{2,2} \left[ x \bigg| \begin{array}{cc}
{(1-\frac{1+\gamma}{\alpha},\frac{1}{\alpha}),(\frac{1}{2}, \frac{1}{2})} \\
{(0,1),(\frac{1}{2},\frac{1}{2})} \end{array}
\right]=\sum_{m=1}^{\infty}Res\{\chi(s)x^{s};s_m \in P(s)\}=\nonumber\\
&=
\sum_{m=0}^{\infty}\lim_{s\rightarrow1+\gamma+m\alpha}\frac{
\left[s-(1+\gamma+m\alpha)\right]\Gamma(s)\Gamma(\frac{1+\gamma-s}{\alpha})}{
\Gamma(\frac{1+s}{2})\Gamma(\frac{1-s}{2})}x^{-s}=
\nonumber\\
&=\frac{\alpha}{\pi}\sum_{m=0}^{\infty} 
\bigg[\frac{(-1)^m\Gamma\left({1+\gamma+m\alpha}\right)}{\Gamma(1+m)}
\cos\left(\frac{
1+\gamma+m\alpha } { 2 }\pi \right)x^{-(1+\gamma+m\alpha)} \bigg],
\end{align}
where we used the limit:
\begin{equation}
\lim_{s\rightarrow1+\gamma+m\alpha}\left[s-(1+\gamma+m\alpha)\right]
\Gamma\left(\frac { 1+\gamma-s }
{\alpha}\right) = \frac{\alpha(-1)^m}{\Gamma(1+m)},
\end{equation}
and the property of $\Gamma(z)$:
\begin{equation}
\Gamma(1-z)\Gamma(z) = \frac{\pi}{\sin({\pi z})},
\label{eq:gammaRelation}
\end{equation}
where here, $z=1/2 + (1+\gamma+m\alpha)/2$.

Applying the ratio test to this series expansion, we get: 
\begin{align}
 \rho_1&=\lim_{m \rightarrow \infty}\bigg|\frac{\Gamma(1+ \gamma
+(m+1)\alpha)\Gamma(m+1)\cos(\pi
\frac{1+\gamma+(m+1)\alpha}{2})}{\Gamma(2+m)\Gamma(1+ \gamma
+m\alpha)\cos(\pi\frac{1+\gamma+m\alpha}{2})}\bigg|\leq\lim_{m \rightarrow
\infty}\bigg|\frac{\Gamma(1+ \gamma
+(m+1)\alpha)\Gamma(m+1)}{\Gamma(2+m)\Gamma(1+ \gamma +m\alpha)}\bigg|=
\nonumber\\
&=\lim_{m \rightarrow \infty}\bigg|\frac{(\gamma +(m+1)\alpha)\Gamma( \gamma
+(m+1)\alpha)}{(m+1)(\gamma+m\alpha)\Gamma(\gamma +m\alpha)}\bigg|=\begin{cases}
0 &  0<\alpha<1  \\
1 &  \alpha=1 \\
\infty &  \alpha>1
\end{cases}
\end{align}
This series expansion converges absolutely for every value of $x \neq 0$ in the
interval $0<\alpha<1$. In this regime of $\alpha$ it is more convenient to write
the $H$-Fox function as a function of $1/x$. Using  Eq. (\ref{eq:iden}) we find:
  
\begin{equation}
 H^{1,1}_{2,2} \left[ x \bigg| \begin{array}{cc}
{(1-\frac{1+\gamma}{\alpha},\frac{1}{\alpha}),(\frac{1}{2}, \frac{1}{2})} \\
{(0,1),(\frac{1}{2},\frac{1}{2})} \end{array} \right]=H^{1,1}_{2,2} \left[
\frac{1}{x} \bigg| \begin{array}{cc} {(1,1),(\frac{1}{2}, \frac{1}{2})} \\
{(\frac{1+\gamma}{\alpha},\frac{1}{\alpha}),(\frac{1}{2},\frac{1}{2})}
\end{array} \right],
\label{eq:TFox2}
\end{equation}.

\item Near $x=0$: The $H$-function is analytic  for $\alpha \in(1,\infty)$
since then $\mu=1-1/\alpha>0$, $\forall x\neq0$. Also for $\alpha=1$, $\mu=0$,
which implies an analytic $H$-Fox function for $-1<x<1$. Choosing the same kind
of contour as above,
this time closing it to the left by a semi-circle of radius $R \rightarrow
\infty$ , we find:
\begin {align}
 H^{1,1}_{2,2} &\left[ x \bigg| \begin{array}{cc}
{(1-\frac{1+\gamma}{\alpha},\frac{1}{\alpha}),(\frac{1}{2}, \frac{1}{2})} \\
{(0,1),(\frac{1}{2},\frac{1}{2})} \end{array}
\right]=-\sum_{m=1}^{\infty}Res\{\chi(s)x^{s};s_m \in Q(s)\}=
\nonumber\\
&=\sum_{m=0}^{\infty}\lim_{s\rightarrow-m}\frac{(s+m)\Gamma(s)\Gamma(\frac{
1+\gamma-s}{\alpha})}{\Gamma(\frac{1+s}{2})\Gamma(\frac{1-s}{2})}x^{-s}
=\nonumber\\
&=\frac{1}
{\pi}\sum_{m=0}^{\infty}\frac{(-1)^m}{ \Gamma(1+m) }
\Gamma\left(\frac{1+\gamma+m}{\alpha}\right)\cos\left(\frac{m}{2}
\pi\right)x^m=
\nonumber\\
&=\frac{1}{\pi}\sum_{m=0}^{\infty}\frac{(-1)^m}{ \Gamma(1+2m) }
\Gamma\left(\frac{1+\gamma+2m}{\alpha}\right)x^{2m}.
\end {align}
where we used the limit:
\begin{equation}
\lim_{s\rightarrow-m}\left(s+m\right)\Gamma(s) =
\frac{(-1)^m}{\Gamma(1+m)},
\end{equation}
and the property of $\Gamma$ in Eq. (\ref{eq:gammaRelation}) for $z=1/2-m/2$.

In this case the ratio test gives:
 \begin{align}
 \rho_2&=\lim_{m \rightarrow \infty}\bigg|\frac{\Gamma(\frac{1+ \gamma
+2(m+1)}{\alpha})\Gamma(2m+1)}{\Gamma(1+2(m+1))\Gamma(\frac{1+ \gamma
+2m}{\alpha})}\bigg|=\nonumber\\
&=\lim_{m \rightarrow \infty}\bigg|\frac{\Gamma(\frac{1+
\gamma +2(m+1)}{\alpha})}{(2m+1)
(2m+2)\Gamma(\frac{1+ \gamma +2m}{\alpha})}\bigg|=
\begin{cases}
\infty &  0<\alpha<1  \\
1 &  \alpha=1 \\
0 &  \alpha>1
\end{cases}
\end{align}   
and the series converges absolutely for every value of $x$ in the
intervals:
\begin{equation}
(-R_1,R_2)=\begin{cases}
(-1,1)& \quad \textit{if} \quad \alpha=1\\
(-\infty,\infty) & \quad \textit{if} \quad \alpha>1
\end{cases}
\end{equation}

\end{enumerate}

\section{$T_{\alpha,\gamma}(x)$ by Weyl Fractional Derivatives} 
\label{app::Fractional}

\subsection{The Weyl Fractional Derivative} 
\label{appsub::weylFrac}

The Weyl fractional derivative of order $\gamma$ of a function $f(x)$,
designated by ${_x}D^{\gamma}_{\infty}$, is defined by
\begin{equation}
({_x}D^{\gamma}_{\infty}f)(x)=(-1)^m\bigg(\frac{d}{dx}\bigg)^m{}_xW^{m-\gamma}_{
\infty}f(x))=(-1)^m\frac{1}{\Gamma(m-\gamma)}\int_{x}^{\infty}\frac{f(t)dt}{
(t-x)^{1+\gamma-m}}\quad \infty<x<\infty
\label{eq:Weyl}
\end{equation} 
where $m-1\leq\gamma<m$, $m\in N$, $\gamma\in C$ and ${}_xW^{\gamma}_{\infty }$ is
the Weyl fractional integral of order $\gamma$ defined by 
\begin{equation}
 ({}_xW_{\gamma}{\infty}f)(x)=\frac{1}{\Gamma(\gamma)}\int_{x}^{\infty}\frac{
f(t)dt}{(t-x)^{1+\gamma-m}}
\end{equation}

\subsection{$T_{\alpha,\gamma}$ by Fractional Derivatives of $L_{\alpha}$ and
$R_{\alpha}$}
\label{appsub::tByFractional}
 
According to Eq. (\ref{eq:Weyl}) the fractional derivative of $\exp(ikx)$ is:
\begin{align}
({_x}D^{\gamma}_{\infty}&[\exp(ikx)])=(-1)^m\bigg(\frac{d}{dx}\bigg)^m{x
} W^{m-\gamma}_{\infty}\exp(ikx)=
\nonumber\\
&=(-1)^m\bigg(\frac{d}{dx}\bigg)^
m(-ik)^{\gamma-m}\exp(ikx)=(-ik)^{\gamma}\exp(ikx).
\end{align}
In the same fashion the Weyl fractional derivative of $\exp(-ikx)$ is: 
\begin{equation}
{_x}D^{\gamma}_{\infty}[\exp(-ikx)]=(-ik)^{\gamma}.
\end{equation}
As a consequence, the Weyl fractional derivative of $\cos(kx)$ and $\sin(kx)$
will be given by:
\begin{equation}
{_x}D^{\gamma}_{\infty}[\cos(kx)]=\cos(kx-\frac{\gamma\pi}{2}),
\end{equation}
and:
\begin{equation}
{_x}D^{\gamma}_{\infty}[\sin(kx)]=\sin(kx-\frac{\gamma\pi}{2}).
\end{equation} 
Using the above definitions we can define the fractional
derivative of order $\gamma$ of $L_{\alpha}$ and $R_{\alpha}$ by:
\begin{align}
{_x}D^{\gamma}_{\infty} [L_{\alpha}(x)]&= 
{_x}D^{\gamma}_{\infty}
\bigg[\frac{1}{\pi}\int_{0}^{\infty}
\cos(kx)\exp(-k^{\alpha})dk\bigg]=\nonumber\\
&=\frac{1}{\pi}\int_{0}^{\infty}({_x}D^{
\alpha}{\infty}[\cos(kx)])(x)\exp(-k^{\alpha})dk=
\nonumber\\
&=\frac{1}{\pi}\int_{0}^{\infty}\cos(kx-\frac{\gamma\pi}{2})\exp(-k^{\alpha})k^{
\gamma}dk,
\end{align}
and:
\begin{align}
{_x}D^{\gamma}_{\infty} [R_{\alpha}(x)]&={_x}D^{\gamma}_{\infty}
\bigg[\frac{1}{\pi}\int_{0}^{\infty}
\sin(kx)\exp(-k^{\alpha})dk]\bigg]=\nonumber\\
&=\frac{1}{\pi}\int_{0}^{\infty}({_x}D^{
\alpha}{\infty}[\sin(kx)])(x)\exp(-k^{\alpha})k^{\gamma}dk=
\nonumber\\
&=\frac{1}{\pi}\int_{0}^{\infty}\sin(kx-\frac{\gamma\pi}{2})\exp(-k^{\alpha})dk.
\end{align} 
By the following identities:
\begin{equation}
\nonumber
\cos(kx-\frac{\gamma\pi}{2})=\cos(kx)
\cos(\frac{\gamma\pi}{2})+\sin(kx)\sin(\frac{\gamma\pi}{2}),
\end{equation}
and
\begin{equation}
 \nonumber
\sin(kx-\frac{\gamma\pi}{2})=\sin(kx)
\cos(\frac{\gamma\pi}{2})-\cos(kx)\sin(\frac{\gamma\pi}{2}),
\end{equation} 
and denoting $\nu_1=\cos(\frac{\gamma\pi}{2})$, and
$\nu_2=\sin(\frac{\gamma\pi}{2})$, we achieve the set of equations:
\begin{align}
&{_x}D^{\gamma}_{\infty}[L_{\alpha}(x)]=\nu_1\frac{1}{\pi}\int_{0}^{\infty}
\cos(kx)\exp(-k^{\alpha})k^{\gamma}dk+\nu_2\frac{1}{\pi}\int_{0}^{\infty}
\sin(kx)\exp(-k^{\alpha})k^{\gamma}dk
\nonumber\\
&{_x}D^{\gamma}_{\infty}
[L_{\alpha}(x)]=\nu_1\frac{1}{\pi}\int_{0}^{\infty}\sin(kx)\exp(-k^{\alpha}
)k^{\gamma}dk-\nu_2\frac{1}{\pi}\int_{0}^{\infty}\cos(kx)\exp(-k^{\alpha})k^{
\gamma}dk.
\label{eq:fractional}
\end{align}
Multiplying the first term by $1/\nu_2$ and the second term by $1/\nu_1$ and
subtracting the first from the second we get:
\begin{equation}
 \frac{1}{\nu_2}L_{\alpha}(x)-\frac{1}{\nu_1}R_{\alpha}(x)=\bigg(\frac{\nu_1}{
\nu_2}+\frac{\nu_2}{\nu_1}\bigg)T_{\alpha,\gamma}(x)
\end{equation} 
which yields:
\begin{equation}
 T_{\alpha,\gamma}=\nu_1 {_x}D^{\gamma}_{\infty}
\left[L_\alpha(x) \right] - \nu_2 {_x}
D^{\gamma}_{\infty}
\left[R_\alpha(x) \right]
\end{equation}

Moreover it can be shown that $T_{\alpha,\gamma}(x)$ is a combination of
fractional derivatives of $H$-Fox functions, since we can represent the
$L_{\alpha}(x)$ and $R_{\alpha}$ by their appropriate $H$-Fox functions. A
well-known result (presented originally by Schneider \cite{schneider1986stable})
gave this representation for the L\'evy $\alpha$-stable distribution. One can
derive it from Eq.(\ref{eq:TFox1}) and Eq.(\ref{eq:TFox2}) by taking $\gamma=0$.
For $0<\alpha<1$: 
\begin{equation}
L_{\alpha}(x) = \frac{1}{\alpha}H^{1,1}_{2,2} \left[x \frac{1}{z} \bigg|
\begin{array}{cc} {(1,1),(\frac{1}{2}, \frac{1}{2})} \\
{(\frac{1}{\alpha},\frac{1}{\alpha}),(\frac{1}{2},\frac{1}{2})} \end{array}
\right] \\, \label{T_ag5_hFox1} 
\end{equation} 
and for $1<\alpha\leq 2$:
 \begin{equation} L_{\alpha}(x) = \frac{1}{\alpha}H^{1,1}_{2,2} \left[ x \bigg|
\begin{array}{cc} {(1-\frac{1}{\alpha},\frac{1}{\alpha}),(\frac{1}{2},
\frac{1}{2})} \\ {(0,1),(\frac{1}{2},\frac{1}{2})} \end{array} \right].
\end{equation}

To represent $R_{\alpha}(x)$ as a $H$-Fox function, we first have to write
$R_{\alpha}(x)$ as a Mellin-Barnes integral: 
\begin{equation}
R_{\alpha}(x)=\frac{1}{\pi}\int_{0}^{\infty}\sin(kx)\exp(-k^{\alpha})dk=-\frac{
1}{\pi}\Im\bigg\{\int_{0}^{\infty}\exp(-ikx)\exp(-k^{\alpha})dk\bigg\}.
\end{equation}
Setting $\exp(-ikx)$ as given in Eq.({\ref{eq:MBexp}}) and integrating over
$k$, we get $R_{\alpha}(x)$:
\begin{equation}
 R_{\alpha}(x)=\frac{1}{2\pi i\alpha}\int_L
\frac{\Gamma(s)\Gamma(\frac{1-s}{\alpha})}{\Gamma(1-\frac{s}{2})\Gamma(\frac{s}{
2})}x^{-s}ds,
\end{equation}
which coincides with the definition of the following $H$-Fox functions.
For $0<\alpha<1$:
\begin{equation}
R_{\alpha}(x) = \frac{1}{\alpha}H^{1,1}_{2,2} \left[x \frac{1}{z} \bigg|
\begin{array}{cc} {(1,1),(1, \frac{1}{2})} \\
{(\frac{1}{\alpha},\frac{1}{\alpha}),(1,\frac{1}{2})} \end{array} \right] \\,
\label{T_ag5_hFox1} \end{equation} 
and for $1<\alpha\leq 2$:
 \begin{equation} R_{\alpha}(x) = \frac{1}{\alpha}H^{1,1}_{2,2} \left[ x \bigg|
\begin{array}{cc} {(1-\frac{1}{\alpha},\frac{1}{\alpha}),(0, \frac{1}{2})} \\
{(0,1),(0,\frac{1}{2})} \end{array} \right] \\. \ 
\end{equation} 
Representing $T_{\alpha,\gamma}(x)$ as fractional derivatives
of $H$-Fox functions gives us a convenient way to calculate Weyl fractional
derivatives of $H$-Fox functions, since a Weyl fractional
derivative of a $H$-Fox is another $H$-Fox function with shifted
indices, given by the relation:
\begin{equation}
 {_x}D^{\alpha}_{\infty}\bigg(x^{\lambda-1}H^{m,n}_{p,q} \left[ x^{\sigma}
\bigg|
\begin{array}{cc} {(a_i,A_i)_{1,p}} \\ {(b_i,B_i)_{1,q}} \end{array}
\right]\bigg )=(-1)^{\Re(\alpha)+1}x^{\lambda-\alpha-1}H^{m+1,n}_{p+1,q+1} \left[
x^{\sigma} \bigg| \begin{array}{cc} {(a_i,A_i)_{1,p},(1-\lambda,\sigma)} \\
{(1-\lambda+\alpha,\sigma),(b_i,B_i)_{1,q}} \end{array} \right],
\end{equation}
 where $\alpha,\lambda\in \mathbb{C}$ and $\Re(\alpha),\sigma>0$
\cite{mathai2009h}.

\bibliographystyle{rsc}
\bibliography{bibliography}

\end{document}